\documentclass[aps,prc,superscriptaddress,showpacs,amsfonts,twocolumn,floatfix]{revtex4-1}

\usepackage{color} %I (Priya) put it in for editing purposes.
\usepackage{amsmath,amssymb} % for using "split", for example, to break the equations 
\usepackage{graphicx}
\usepackage{lineno}
\definecolor{gray01}{gray}{0.9}
\definecolor{gray02}{gray}{0.8}
\definecolor{gray03}{gray}{0.7}
\definecolor{gray04}{gray}{0.6}
\definecolor{gray05}{gray}{0.5}
\definecolor{gray06}{gray}{0.4}
\definecolor{gray07}{gray}{0.3}
\definecolor{gray08}{gray}{0.2}
\definecolor{gray09}{gray}{0.1}

\newcommand{\gr}[1]{{\color{green} #1}}
\newcommand{\bl}[1]{{\color{blue} #1}}
\newcommand{\re}[1]{{\color{red} #1}}
\newcommand{\ma}[1]{{\color{magenta} #1}}

\begin{document}

\title{Measurement of single- and double-polarization observables in the photoproduction of 
$\pi^+\pi^-$~meson pairs off the proton using CLAS at Jefferson Laboratory}

%%%%%%%%%%%%%%% Latex Macros for institute addresses  %%%%%%%%%%%%%%%%%%%%%%%%% 
%\newcommand*{\FSU}{Florida State University, Tallahassee, Florida 32306}
%\newcommand*{\FSUindex}{13}
%\affiliation{\FSU}
%\newcommand*{\BONN}{Helmholtz-Institut f\"ur Strahlen- und Kernphysik, Universit\"at Bonn, 53115 Bonn, Germany}
%\affiliation{\BONN}
%\newcommand*{\ITEP}{Institute of Theoretical and Experimental Physics, Moscow, 117259, Russia}
%\newcommand*{\ITEPindex}{22}
%\affiliation{\ITEP}
%\newcommand*{\JINR}{Joint Institute for Nuclear Research, 141980 Dubna, Russia}
%\affiliation{\JINR}
%\newcommand*{\MSU}{Skobeltsyn Institute of Nuclear Physics, Lomonosov Moscow State University, 119234 Moscow, Russia}
%\newcommand*{\MSUindex}{32}
%\affiliation{\MSU}
%\newcommand*{\JLAB}{Thomas Jefferson National Accelerator Facility, Newport News, Virginia 23606}
%\newcommand*{\JLABindex}{35}
%\affiliation{\JLAB}
%\newcommand*{\NRC}{NRC ``Kurchatov Institute", PNPI, 188300, Gatchina, Russia}
%affiliation{\NRC}

\newcommand*{\ANL}{Argonne National Laboratory, Argonne, Illinois 60439, USA}
\newcommand*{\ANLindex}{1}
\affiliation{\ANL}
\newcommand*{\ASU}{Arizona State University, Tempe, Arizona 85287-1504, USA}
\newcommand*{\ASUindex}{2}
\affiliation{\ASU}
\newcommand*{\CSU}{California State University, Dominguez Hills, Carson, California 90747, USA}
\affiliation{\CSU}
\newcommand*{\CANISIUS}{Canisius University, Buffalo, New York, USA}
\newcommand*{\CANISIUSindex}{3}
\affiliation{\CANISIUS}
\newcommand*{\CMU}{Carnegie Mellon University, Pittsburgh, Pennsylvania 15213, USA}
\newcommand*{\CMUindex}{4}
\affiliation{\CMU}
\newcommand*{\CUA}{Catholic University of America, Washington, D.C. 20064, USA}
\newcommand*{\CUAindex}{5}
\affiliation{\CUA}
\newcommand*{\SACLAY}{IRFU, CEA, Universit\'e Paris-Saclay, F-91191 Gif-sur-Yvette, France}
\newcommand*{\SACLAYindex}{6}
\affiliation{\SACLAY}
\newcommand*{\CNU}{Christopher Newport University, Newport News, Virginia 23606, USA}
\newcommand*{\CNUindex}{7}
\affiliation{\CNU}
\newcommand*{\UCONN}{University of Connecticut, Storrs, Connecticut 06269, USA}
\newcommand*{\UCONNindex}{8}
\affiliation{\UCONN}
\newcommand*{\DUKE}{Duke University, Durham, North Carolina 27708-0305, USA}
\newcommand*{\DUKEindex}{9}
\affiliation{\DUKE}
\newcommand*{\DUQUESNE}{Duquesne University, 600 Forbes Avenue, Pittsburgh, Pennsylvania 15282, USA}
\newcommand*{\DUQUESNEindex}{10}
\affiliation{\DUQUESNE}
\newcommand*{\FU}{Fairfield University, Fairfield, Connecticut 06824, USA}
\newcommand*{\FUindex}{11}
\affiliation{\FU}
\newcommand*{\FERRARAU}{Universit\`a di Ferrara, 44121 Ferrara, Italy}
\newcommand*{\FERRARAUindex}{12}
\affiliation{\FERRARAU}
\newcommand*{\FIU}{Florida International University, Miami, Florida 33199, USA}
\affiliation{\FIU}
\newcommand*{\FSU}{Florida State University, Tallahassee, Florida 32306, USA}
\newcommand*{\FSUindex}{13}
\affiliation{\FSU}
\newcommand*{\GWUI}{The George Washington University, Washington, D.C. 20052, USA}
\newcommand*{\GWUIindex}{14}
\affiliation{\GWUI}
\newcommand*{\GSIFFN}{GSI Helmholtzzentrum f\"ur Schwerionenforschung GmbH, D-64291 Darmstadt, Germany}
\newcommand*{\GSIFFNindex}{15}
\affiliation{\GSIFFN}
\newcommand*{\BONN}{Helmholtz-Institut f\"ur Strahlen- und Kernphysik, Universit\"at Bonn, 53115 Bonn, Germany}
\affiliation{\BONN}
\newcommand*{\INFNCA}{INFN, Sezione di Catania, 95123 Catania, Italy}
\affiliation{\INFNCA}
\newcommand*{\INFNFE}{INFN, Sezione di Ferrara, 44100 Ferrara, Italy}
\newcommand*{\INFNFEindex}{16}
\affiliation{\INFNFE}
\newcommand*{\INFNFR}{INFN, Laboratori Nazionali di Frascati, 00044 Frascati, Italy}
\newcommand*{\INFNFRindex}{17}
\affiliation{\INFNFR}
\newcommand*{\INFNGE}{INFN, Sezione di Genova, 16146 Genova, Italy}
\newcommand*{\INFNGEindex}{18}
\affiliation{\INFNGE}
\newcommand*{\INFNRO}{INFN, Sezione di Roma Tor Vergata, 00133 Rome, Italy}
\newcommand*{\INFNROindex}{19}
\affiliation{\INFNRO}
\newcommand*{\INFNTUR}{INFN, Sezione di Torino, 10125 Torino, Italy}
\newcommand*{\INFNTURindex}{20}
\affiliation{\INFNTUR}
\newcommand*{\INFNPAV}{INFN, Sezione di Pavia, 27100 Pavia, Italy}
\newcommand*{\INFNPAVindex}{21}
\affiliation{\INFNPAV}
\newcommand*{\NRC}{NRC ``Kurchatov Institute", PNPI, 188300, Gatchina, Russia}
\affiliation{\NRC}
\newcommand*{\ORSAY}{Universit\'e Paris-Saclay, CNRS/IN2P3, IJCLab, 91405 Orsay, France}
\newcommand*{\ORSAYindex}{22}
\affiliation{\ORSAY}
\newcommand*{\JMU}{James Madison University, Harrisonburg, Virginia 22807, USA}
\newcommand*{\JMUindex}{23}
\affiliation{\JMU}
\newcommand*{\KNU}{Kyungpook National University, Daegu 41566, Republic of Korea}
\newcommand*{\KNUindex}{24}
\affiliation{\KNU}
\newcommand*{\LAMAR}{Lamar University, 4400 MLK Blvd, PO Box 10046, Beaumont, Texas 77710, USA}
\newcommand*{\LAMARindex}{25}
\affiliation{\LAMAR}
\newcommand*{\MIT}{Massachusetts Institute of Technology, Cambridge, Massachusetts 02139-4307, USA}
\newcommand*{\MITindex}{26}
\affiliation{\MIT}
\newcommand*{\MISS}{Mississippi State University, Mississippi State, MS 39762-5167, USA}
\newcommand*{\MISSindex}{27}
\affiliation{\MISS}
\newcommand*{\UNH}{University of New Hampshire, Durham, New Hampshire 03824-3568, USA}
\newcommand*{\UNHindex}{28}
\affiliation{\UNH}
\newcommand*{\NMSU}{New Mexico State University, PO Box 30001, Las Cruces, New Mexico 88003, USA}
\newcommand*{\NMSUindex}{29}
\affiliation{\NMSU}
\newcommand*{\NSU}{Norfolk State University, Norfolk, Virginia 23504, USA}
\newcommand*{\NSUindex}{30}
\affiliation{\NSU}
\newcommand*{\OHIOU}{Ohio University, Athens, Ohio 45701, USA}
\newcommand*{\OHIOUindex}{31}
\affiliation{\OHIOU}
\newcommand*{\ODU}{Old Dominion University, Norfolk, Virginia 23529, USA}
\newcommand*{\ODUindex}{32}
\affiliation{\ODU}
\newcommand*{\JLUGiessen}{II. Physikalisches Institut der Universit\"at Giessen, 35392 Giessen, Germany}
\newcommand*{\JLUGiessenindex}{33}
\affiliation{\JLUGiessen}
\newcommand*{\ROMAII}{Universit\`a di Roma Tor Vergata, 00133 Rome, Italy}
\newcommand*{\ROMAIIindex}{34}
\affiliation{\ROMAII}
\newcommand*{\SDU}{Shandong University, Qingdao, Shandong 266237, China}
\newcommand*{\SDUindex}{35}
\affiliation{\SDU}
\newcommand*{\MSU}{Skobeltsyn Institute of Nuclear Physics, Lomonosov Moscow State University, 119234 Moscow, Russia}
\newcommand*{\MSUindex}{36}
\affiliation{\MSU}
\newcommand*{\SCAROLINA}{University of South Carolina, Columbia, South Carolina 29208, USA}
\newcommand*{\SCAROLINAindex}{37}
\affiliation{\SCAROLINA}
\newcommand*{\TEMPLE}{Temple University, Philadelphia, Pennsylvania 19122, USA}
\newcommand*{\TEMPLEindex}{38}
\affiliation{\TEMPLE}
\newcommand*{\JLAB}{Thomas Jefferson National Accelerator Facility, Newport News, Virginia 23606, USA}
\newcommand*{\JLABindex}{39}
\affiliation{\JLAB}
\newcommand*{\UTFSM}{Universidad T\'{e}cnica Federico Santa Mar\'{i}a, Casilla 110-V Valpara\'{i}so, Chile}
\newcommand*{\UTFSMindex}{40}
\affiliation{\UTFSM}
\newcommand*{\BRESCIA}{Universit\`a degli Studi di Brescia, 25123 Brescia, Italy}
\newcommand*{\BRESCIAindex}{41}
\affiliation{\BRESCIA}
\newcommand*{\UCR}{University of California Riverside, 900 University Avenue, Riverside, California 92521, USA}
\newcommand*{\UCRindex}{42}
\affiliation{\UCR}
\newcommand*{\GLASGOW}{University of Glasgow, Glasgow G12 8QQ, United Kingdom}
\newcommand*{\GLASGOWindex}{43}
\affiliation{\GLASGOW}
\newcommand*{\YORK}{University of York, York YO10 5DD, United Kingdom}
\newcommand*{\YORKindex}{44}
\affiliation{\YORK}
\newcommand*{\VT}{Virginia Tech, Blacksburg, Virginia 24061-0435, USA}
\newcommand*{\VTindex}{45}
\affiliation{\VT}
\newcommand*{\VIRGINIA}{University of Virginia, Charlottesville, Virginia 22901, USA}
\newcommand*{\VIRGINIAindex}{46}
\affiliation{\VIRGINIA}
\newcommand*{\WM}{College of William and Mary, Williamsburg, Virginia 23187-8795, USA}
\newcommand*{\WMindex}{47}
\affiliation{\WM}
\newcommand*{\YEREVAN}{Yerevan Physics Institute, 375036 Yerevan, Armenia}
\newcommand*{\YEREVANindex}{48}
\affiliation{\YEREVAN}

\newcommand*{\NOWJLAB}{Thomas Jefferson National Accelerator Facility, Newport News, Virginia 23606, USA}
\newcommand*{\NOWISU}{Idaho State University, Pocatello, Idaho 83209, USA}
\newcommand*{\NOWANL}{Argonne National Laboratory, Argonne, Illinois 60439, USA}

%%%%%%%%%%%%%%% END OF Latex Macros for institute addresses  %%%%%%%%%%%%%%%%%%%%%%%%% 


\author{P.~Roy} \affiliation{\FSU}
\author{S.~Cao} \affiliation{\FSU}
\author{V.~Crede} \altaffiliation[Corresponding author: crede@fsu.edu]{}\affiliation{\FSU}
\author{E.~Klempt} \affiliation{\BONN} \affiliation{\JLAB}
\author{V.~A.~Nikonov} \affiliation{\BONN} 
\author{A.~V.~Sarantsev} \affiliation{\BONN} \affiliation{\NRC}
\author {V.~D.~Burkert} \affiliation{\JLAB}
%\author {E.~Golovatch} 
%\affiliation{\MSU}
%\author {C.~D.~Keith}
%\affiliation{\JLAB}
\author {V.~Mokeev} \affiliation{\JLAB}
%\affiliation{\JLAB}

\author {P.~Achenbach} 
\affiliation{\JLAB}
\author {J.~S.~Alvarado} 
\affiliation{\ORSAY}
% ...

%\thanks{}
%\affiliation{}

\collaboration{The CLAS Collaboration}
\noaffiliation

\date{Received: \today / Revised version:}

\begin{abstract}
The photoproduction of $\pi^+\pi^-$~meson pairs off the proton has been studied in the reaction $\gamma p 
\to p\,\pi^+\pi^-$ using the CEBAF Large Acceptance Spectrometer (CLAS) and the frozen-spin target (FROST)
in Hall~B at the Thomas Jefferson National Accelerator Facility. For the first time, the beam and target asymmetries,
$I^{s,c}$ and $P_{x,y}$, have been measured along with the beam-target double-polarization observables, $P^{s,c}_{x,y}$,
using a transversely polarized target with center-of-mass energies ranging from 1.51~GeV up to 2.04~GeV. These data
and additional $\pi\pi$~photoproduction observables from CLAS and experiments elsewhere were included in a
partial-wave analysis within the Bonn-Gatchina framework. Significant contributions from $s$-channel resonance
production are observed in addition to $t$-channel exchange processes. The data indicate significant contributions
from $N^\ast$ and $\Delta^\ast$~resonances in the third and fourth resonance regions.
\end{abstract}

\maketitle
%\tableofcontents

\section{Introduction}
The study of two-pion photoproduction provides insight into the rich dynamics of quantum chromodynamics. 
At higher energies, inelastic diffractive $\rho^0$~photoproduction provides the largest contribution to the 
photoproduced $\pi^+\pi^-$ cross section. The integrated cross section is found to rise slowly with energy,
reaching $\approx 3~\mu$b at $E_\gamma = 20$~GeV and $\approx 4~\mu$b at 100~GeV. The process can be 
understood in terms of the incident photon fluctuating into a $\rho^0$~vector meson with the same $J^{PC}$~quantum
numbers as the photon and then scattering off the proton by Pomeron or Reggeon exchanges~\cite{H1:2009cml}. The
diffractively produced vector meson can be considered the hadronic analogue of the photon. Or alternatively, the photon
can be considered as a superposition of the lightest vector mesons, which is the basis for the Vector Meson Dominance
Model. The hadronic composition of the photon allows for a strong interaction between the photon and the target nucleon,
with a composite electrically neutral object acting as a mediator. The mechanism of diffractive $\rho^0$~production is
thus fundamentally different from deep-inelastic scattering, and understanding this process in terms of the underlying
quark-gluon structure is still a major challenge for quantum chromodynamics. A comprehensive discussion of this topic
can be found in~Ref.~\cite{Brodsky:1994kf}. 

At lower energies, the production of baryon resonances is more relevant. The $\Delta(1232)$~resonance becomes one of
the dominant contributors as well as higher-mass $N^\star$~resonances and $\Delta$~excitations. The photoproduction
of pion pairs is important in the search for hitherto unknown nucleon and $\Delta$~states that have very small couplings
to the $p\pi$~ground state. Such excited quark configurations may de-excite through a cascading process by populating
intermediate states involving $N$ and $\Delta$~excitations as intermediate isobars, which still carry angular momentum
and can further decay by emitting an additional pion, resulting in an $N\pi\pi$~final state. The study of such decay
cascades has led to the identification of a new nucleon resonance, $N(1880)\,1/2^+$. This is particularly interesting since
the four states, $N(1880)\,1/2^+$, $N(1900)\,3/2^+$, $N(2000)\,5/2^+$, and $N(1990)\,7/2^+$, are now considered to
form a previously missing quartet of nucleon resonances with spin~$\frac{3}{2}$~\cite{Anisovich:2011su,
CBELSATAPS:2015kka,CBELSATAPS:2015taz}. This group of resonances has previously been predicted by traditional quark
models based on three symmetric quark degrees of freedom. However, the identification of these resonances is
incompatible with the static diquark-quark picture of the nucleon, with excitations in the diquark frozen out, since the
$({\bf 70},\,2_2^+)$~multiplet requires excitations between all three valence quarks. Recent reviews of the progress
toward understanding the baryon spectrum are available in Refs.~\cite{Klempt:2009pi,Crede:2013kia,Ireland:2019uwn,
Thiel:2022xtb,Gross:2022hyw}, for instance.

The decay of baryon resonances into vector mesons remains largely underexplored and has experimentally mostly
focused on the $\omega$~meson. In contrast, baryon decays into $\rho$~mesons  are fairly unknown. The large width
of $\Gamma_\rho \approx 150$~MeV makes the preparation of a pure $\gamma p\to p\rho^0\to p\pi^+\pi^-$ event
sample almost impossible and therefore, the $\rho$~decays need to be analyzed as part of a broader
$p\pi^+\pi^-$~analysis, which includes $p\pi^+$, $p\pi^-$, or other $\pi^+\pi^-$~contributions. Such an approach
is substantially more challenging in an amplitude analysis. Based on the classification scheme in the Review of Particle
Physics~\cite{ParticleDataGroup:2022pth}, the evidence of existence for a nucleon or $\Delta$~resonance is either
non-existent or poor (one-star assignment) in $\rho$~decays.

In this paper, first measurements are presented for the target asymmetries $\rm\bf P_x$ and $\rm\bf P_y$, the linearly
polarized beam asymmetries $\rm\bf I^s$ and $\rm\bf I^c$, as well as results for the four beam-target double-polarization 
asymmetries $\rm\bf P_x^{\,s,c}$ and $\rm\bf P_y^{\,s,c}$, in the photon-induced reaction off the proton:
\begin{equation}
\gamma\,p\,\to\,p\,\pi^+\pi^-,
\label{Equation:two_pion_reaction}
\end{equation} 
from the CLAS-g9b (FROST) experiment. These new measurements cover a broad range in incident-photon energies,
$E_{\gamma} \in [0.7, 1.8]$~GeV, which translates into a center-of-mass energy range of about $W\in [1.51, 2.04]$~GeV. 
This paper has the following structure. Section~\ref{Section:ExperimentalSetup} describes the relevant components
of the CLAS-g9b (FROST) experimental setup. The data reconstruction and event selection are  discussed in
Section~\ref{Section:Selection} and the technique for extracting the polarization observables is described in
Section~\ref{Section:Observables}. Experimental results and a discussion of the observed resonance contributions
are presented in Sections~\ref{Section:Results} and \ref{Section:PWA}, respectively. The paper ends with a brief
summary and an outlook.

\subsection*{Previous Measurements\label{ssec:PreviousMeasurements}}
The production of two pions off the nucleon in the resonance regime has been studied in $\pi^-$-induced reactions
and using electromagnetic probes. A more detailed summary of these measurements and relevant physics conclusions
can be found in our companion publication from the CLAS Collaboration on the photoproduction of two charged pions
off protons~\cite{CLAS:2024iir} and references therein. In a very brief summary, studies after 2000 on
$\pi^- p\to N\pi\pi$ were reported by the Crystal Ball Collaboration at BNL for the 
$\pi^0\pi^0\,n$ final state~\cite{CrystalBall:2004qln} and, more recently, by the HADES Collaboration at GSI for the
$\pi^+\pi^-\,n$ as well as $\pi^0\pi^- p$ final states~\cite{HADES:2020kce}. Results on $\pi\pi N$~photoproduction
cross sections can be found in Ref.~\cite{CLAS:2024iir}, and further references therein.

Polarization observables in photoproduction, their definitions and extraction from data, are discussed in
Section~\ref{Section:Observables}. Here, a brief summary of the experimental references is given. The data on
polarization observables in $\gamma p\to N\pi\pi$ have been recorded for the most part at the MAinz MIcrotron
(MAMI)~\cite{Kaiser:2008zza} and the ELectron Stretcher Accelerator (ELSA) facility~\cite{Hillert:2006yb}. The
A2~Collaboration at MAMI (together with other collaborations) has extracted the beam-helicity asymmetry
${\bf I^{\,\odot}}$ for $\pi^0\pi^0$ and $\pi^+\pi^0$ production off the nucleon~\cite{Zehr:2012tj}, the target
asymmetries ${\bf P_{\,x,\,y}}$ (transverse target polarization) as well as the double-polarization observables
${\bf P^{\,\odot}_{\,x,\,y}}$ (circular beam and transverse target polarization)~\cite{A2:2022ipx}, and the helicity
dependent ${\bf E}$~observable (circular beam and longitudinal target polarization) for the $p\pi^0\pi^0$ final
state~\cite{GDH:2005jgl}. The helicity dependence was also measured for the $p\pi^+\pi^-$~final
state~\cite{GDH:2007nkn}.

\begin{figure}[!t]
  \includegraphics[width=0.41\textwidth,height=0.26\textheight]{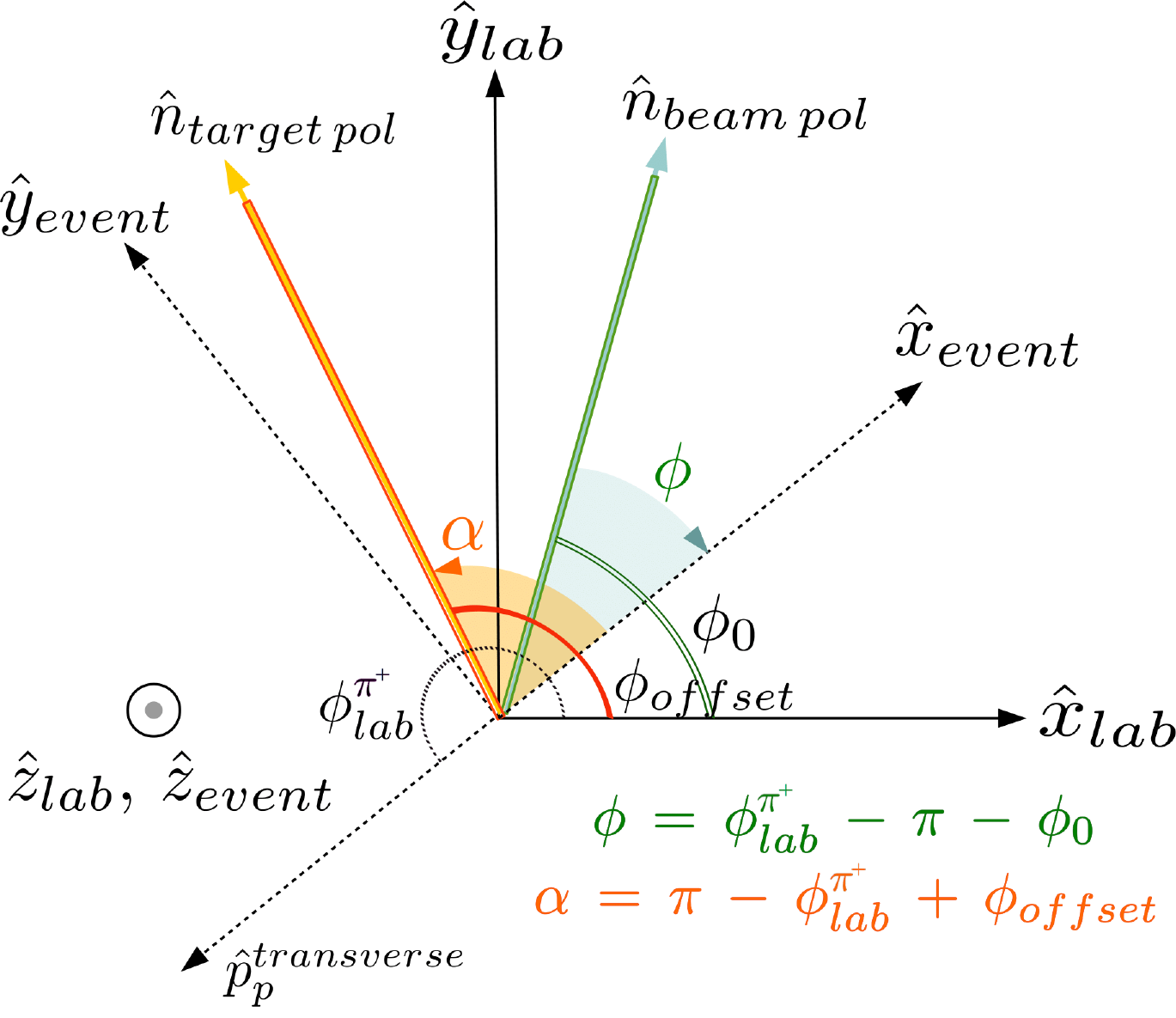}
  \caption{An illustration of relevant coordinate systems, and the directions of polarization of the linearly
    polarized photon beam and the transversely polarized butanol target in the laboratory and event frames.
    See text for more information on the definition of the axes. The beam polarization (shown as the green-shaded
    area) was inclined at an angle $\phi_0=0^\circ$ with respect to the $x$-axis in the laboratory frame
    ($\hat{x}_{\rm \,lab}$) for the parallel setting and at $\phi_0=90^{\circ}$ for the perpendicular setting. The target
    polarization (shown as the orange-shaded area) was inclined at an angle $\phi_{\rm \,offset}$ with respect to the
    $\hat{x}_{\rm \,lab}$-axis. Also shown in the sketch are the azimuthal angle $\phi$ ($\alpha$) of the beam (target)
    polarization in the event frame and its relation with the azimuthal angle $\phi^{\,\pi^+}_{\rm \,lab}$ of the recoiling
    $\pi^+$~meson in the laboratory frame. More details on how these angles were used in the analysis are discussed
    in Section~\ref{Section:Observables}. Reprinted figure with permission from~\cite{CLAS:2017jrx}, Copyright (2018)
    by the American Physical Society.}\label{Figure:axes_and_angles_def}
\end{figure}

The CBELSA/TAPS Collaboration at ELSA reported on various polarization observables for the photoproduction of
$\pi^0$~pairs including the beam asymmetries ${\bf I^{\,s,c}}$~\cite{CBELSATAPS:2015tyg} and, using a transversely
polarized target, the target asymmetries ${\bf P_{\,x,\,y}}$ as well as the double-polarization observables
${\bf P^{\,s,c}_{\,x,y}}$~\cite{CBELSATAPS:2022uad}. Two additional measurements complete the landscape of
polarization observables. The CLAS Collaboration at JLab studied the beam helicity asymmetry in the reaction
$\gamma p\to p\pi^+\pi^-$~\cite{CLAS:2005oqk} and the LEPS Collaboration reported on the  photon-beam
asymmetry in the reaction $\gamma p\to \pi^-\Delta^{++}$ at forward angles of the produced
$\pi^-$~mesons~\cite{LEPS:2018pbi}. 

The polarization observables ${\bf I^{\,s,c}}$, ${\bf P_{\,x,y}}$, and ${\bf P^{\,s,c}_{\,x,y}}$ reported here are first
measurements for the photoproduced $p\,\pi^+\pi^-$ final state involving two charged pions.

\section{The FROST Experimental Setup\label{Section:ExperimentalSetup}}
The FROzen Spin Target (FROST)~\cite{Keith:2012ad} experiment was conducted at the Thomas Jefferson National
Accelerator Facility (Jefferson Lab) in Newport News, Virginia, using the CEBAF Large Acceptance Spectrometer
(CLAS)~\cite{Mecking:2003zu} in Hall B at Jefferson Lab. The FROST (CLAS-g9) experiment consisted of a variety
of individual experiments combining linear and circular beam polarization with longitudinal (CLAS-g9a) and transverse
(CLAS-g9b) target polarization, thus providing access to single- and double-polarization observables in a large number
of photoproduction reactions~\cite{CLAS:2005oqk,CLAS:2015pjm,CLAS:2017yjv,CLAS:2017jrx,CLAS:2018mmb}. For the
$\pi^+\pi^-$~measurements presented here, the target was transversely polarized and the beam was linearly polarized.
Figure~\ref{Figure:axes_and_angles_def} shows a schematic that illustrates the complex kinematic situation of linear
beam polarization in combination with transverse target polarization in two coordinate systems relevant for this analysis:
the laboratory frame and the event frame. In both frames, the $z$-axis was chosen to be along the direction of the
incoming photon beam. The positive direction of the $y$-axis in the laboratory frame, $\hat{y}_{\rm \,lab}$, was defined
as the vertical direction pointing away from the floor of the experimental hall, and $\hat{x}_{\rm \,lab}$ was then given
by $\hat{x}_{\rm \,lab} = \hat{y}_{\rm \,lab}\,\times\,\hat{z}_{\rm \,lab}$. The $x$- and $y$-axes in the event frame were
chosen in the following way: $\hat{y}_{\rm \,event}$ was perpendicular to the center-of-mass production plane spanned
by the photon-beam axis and the recoiling proton. Mathematically, $\hat{y}_{\rm \,event} =
(\hat{p}_{\,p}\,\times\,\hat{z}_{\rm \,event}) / {\left| \hat{p}_{\,p}\,\times\,\hat{z}_{\rm \,event}\right|}$, where $\hat{p}_{\,p}$
is a unit vector along the momentum of the recoiling proton in the center-of-mass frame. Then, $\hat{x}_{\rm \,event}$
was given by $\hat{x}_{\rm \,event} = \hat{y}_{\rm \,event} \times \hat{z}_{\rm \,event}$.

The coherent bremsstrahlung technique was used to prepare the incident beam of linearly polarized tagged photons.
This technique is based on unpolarized electrons scattering off a crystal lattice which is oriented such that the desired
lattice vector is nearly perpendicular to the incident electron beam. As a result, the bremsstrahlung cross section is
greatly enhanced at certain values of the momentum transfer {\bf q}. By varying the crystal orientation, these enhancements
or peaks can be moved to provide the greatest photon flux at the energy of interest. Moreover, a high degree of linear
polarization of the outgoing photons is observed in the plane of~{\bf q} and the incident beam. A detailed description
of this technique is given in Refs.~\cite{Timm:1969mf,Lohmann:1994vz}. For the measurements presented here,
unpolarized electrons with energies up to 5.5~GeV from the Continuous Electron Beam Accelerator Facility (CEBAF) at
Jefferson Lab scattered off a $50~{\rm \mu m}$~thick diamond radiator and were deflected into the CLAS tagging
system~\cite{Sober:2000we}. This detector system consisted of a layer of 384 partially overlapping small scintillators
on top of a layer of 61~scintillator paddles, and both layers combined provided the relevant information for tagging the
time and the energy of the corresponding bremsstrahlung photons with a resolution of $\Delta t\sim 100$~ps and
$\Delta E/E\approx 0.1\,\%$, respectively. The orientation of the linear polarization could be set to either parallel 
(in the following, denoted as $\parallel$) or to perpendicular (denoted as $\perp$) relative to the floor of the 
experimental hall by adjusting the azimuthal angle of the crystal lattice of the diamond radiator~\cite{Livingston:2008hv}.
The corresponding azimuthal angle of the beam polarization in the laboratory frame was then $\phi_0 = 0^{\circ}$ or
$90^{\circ}$, respectively (see Fig.~\ref{Figure:axes_and_angles_def}). The angle between a selected diamond plane and
the incident electron beam determined the leading-edge of the enhancement in the photon energy spectrum, known as
the ``coherent edge.'' A high degree of linear polarization is obtained within a window of about $200$~MeV below the
coherent edge. In this experiment, the coherent edges covered a photon energy range from $0.9$~GeV to $2.1$~GeV
in intervals of $0.2$~GeV. The average degree of polarization was determined via a fit of the photon energy spectrum
using a coherent bremsstrahlung calculation~\cite{Ken_pol_cal} and was found to vary between 40--60\,\% depending
on energy.

\begin{figure}[t]
  \includegraphics[height=0.26\textheight,width=0.45\textwidth]{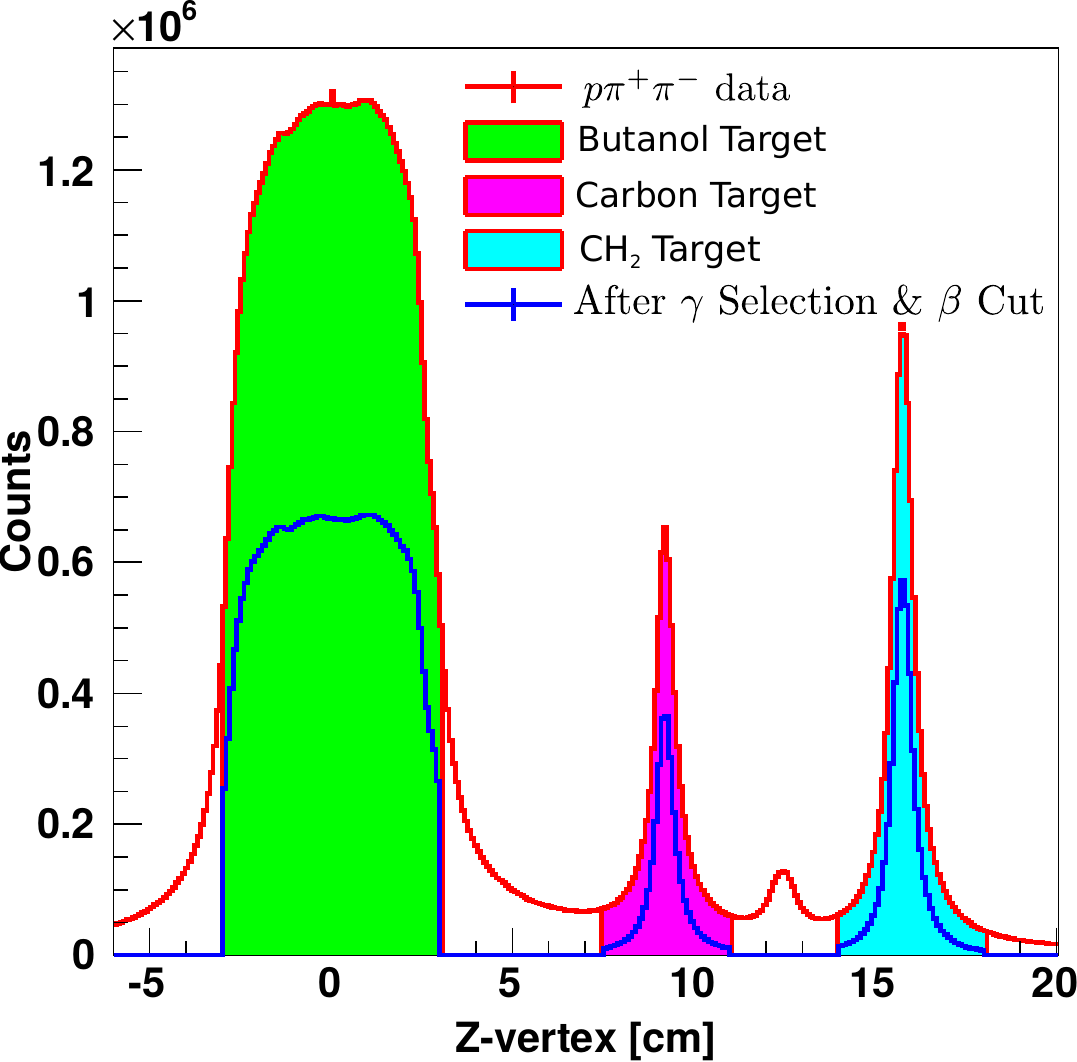}
  \caption{The $z$-vertex distribution (axis along the beamline) of all reconstructed particles. The CLAS center was
    chosen as the $z=0$~coordinate. The peak on the left shows the $z$~position of the butanol target, the peak
    situated next to it shows the position of the carbon target and the peak on the right shows the position of the
    polyethylene (CH$_2$) target. The \re{red} line denotes the data containing all $p\,\pi^+\pi^-$ events. The \bl{blue}
    line denotes these events after applying photon selection and particle-identification cuts (discussed in 
    Section~\ref{Section:Selection}). The small peak between the carbon and the polyethylene target originated from the
    end-cap of the heat shield. Reprinted figure with permission from~\cite{CLAS:2017jrx}, Copyright (2018) by the
    American Physical Society.}\label{fig:Vertex_cut}
\end{figure}

Protons in TEMPO-doped butanol (C$_4$H$_9$OH) served as the target nucleons. The target material was contained
in the $5$~cm long frozen-spin butanol target system~\cite{Keith:2012ad} and was transversely polarized using a
Dynamic Nuclear Polarization (DNP) technique~\cite{Abragam:1978}. At CLAS, this technique of dynamically polarizing
nucleons by continuous microwave irradiation at 140 GHz was also used earlier with electron beams up to 10~nA and
has been previously described in Ref.~\cite{Keith:2003ca}. In the FROST experiment, the target was initially polarized
outside the CLAS detector in a homogeneous magnetic field of 5~T and at a temperature of $T=200$--300~mK. The
target was then cooled down to about 60~mK and a 0.5~T holding field using a dipole magnet was applied to maintain
the transverse polarization inside the CLAS detector system. An average transverse polarization of about $81\,\%$ was
achieved. The degree of polarization was determined based on regular NMR measurements taken for both target
polarizations: pointing away from the floor (denoted as `$+$') and pointing towards the floor (denoted as `$-$').
Moreover, the target polarization was inclined at an angle $\phi_{\rm \,offset}\,=\,116.1^\circ\,\pm\,0.4^{\circ}$
(referred to as the target offset angle) from the $x$-axis in the laboratory frame for the `$+$' setting and at 
$\phi_{\rm \,offset}\,=\,-63.9^\circ\,\pm\,0.4^{\circ}$ for the `$-$' setting, as shown in 
Fig.~\ref{Figure:axes_and_angles_def}. These offsets were necessary to prevent photoproduced $e^+e^-$~background
from being directed into the CLAS acceptance region by the target holding field. 

In addition to the main butanol target, two unpolarized targets were placed in the cryostat. A carbon target was used
for background studies and a polyethylene (CH$_2$) target for additional systematic studies involving unpolarized
nucleons. These were placed further downstream from the butanol target at approximately $\Delta z = 9$~cm and 16~cm, 
respectively, and were well-separated from each other, as is evident from the $z$-vertex distribution shown in
Fig.~\ref{fig:Vertex_cut}. The thickness of the additional targets was chosen such that the hadronic rate from each
was about 10\,\% of the rate for butanol.

The charged final-state particles were detected using the CLAS spectrometer~\cite{Mecking:2003zu} based on a toroidal
magnet that provides a non-homogeneous magnetic field, primarily pointing in the $\phi$~direction, with a maximum
magnitude of $1.8$~T generated by a six-coil torus magnet. The design of the magnet provided a field-free region
around the polarized target. The CLAS detector system had many components, each with a six-fold symmetry about
the beam axis, covering a solid angle of about 80\,\% of~$4\pi$. For an event to be recorded, the trigger configuration
required the detection of at least one charged track.

In a brief summary of the spectrometer, the target cell was surrounded by the start counter~\cite{Sharabian:2005kq},
which was divided into six sectors, each containing four scintillator paddles. It was used to determine the vertex time
of the events. After traversing the start counter, the charged final-state particles were deflected by the toroidal magnetic
field and the tracks were measured by a set of three drift-chamber packages~\cite{Mestayer:2000we}, which provided a
polar angle coverage of $8^{\circ}$ to $142^{\circ}$ and an azimuthal angle coverage of about 80\,\%. From the curvature
in the magnetic field, the particle momenta were determined with a momentum and angle resolution of 
$\Delta p/p\approx 1\,\%$ and $\Delta\theta\approx 1$\,-\,$2^\circ$, respectively. Finally, the time-of-flight was
measured by a set of 288~scintillator paddles with a timing resolution of 80~ps in short counters and 160~ps in long
counters~\cite{Smith:1999ii}. The vertex-information from the start counter together with the time-of-flight information
allowed the determination of the velocity of the charged particles. The momentum and the velocity information combined
allowed a mass determination and aided in the particle identification. 

\section{Event Selection\label{Section:Selection}}
The preparation of the $\gamma p\to p\pi^+ \pi^-$ event sample was similar to the procedure outlined in 
Ref.~\cite{CLAS:2017jrx} for the $p\omega\to p\pi^+ \pi^- \pi^0$ final state involving a missing~$\pi^0$. Events for
the charged double-pion reaction were either reconstructed fully exclusively (Topology 4) or one missing particle could
be identified and the event fully reconstructed from the overconstrained event kinematics. In the following, a reaction
involving an undetected $\pi^+$ $(\pi^-)$ is referred to as Topology~1~(2) and involving an undetected proton as
Topology~3. The fully exclusive reaction was the basis for the determination of kinematic corrections in both reactions,
$\gamma p\to p\pi^+\pi^-$ (discussed here) and $\gamma p\to p\omega\to p\pi^+\pi^-\pi^0$~\cite{CLAS:2017jrx},
and the study of systematics. Events for Topology~4 were selected when exactly one final-state proton as well as one
$\pi^+$ and one $\pi^-$~track could be identified in the event reconstruction. A four-constraint (4C) kinematic fit
imposing energy and momentum conservation was then used in Topology~4 and a one-constraint (1C) kinematic fit
was used in Topologies 1-3 to reconstruct the missing particle.

\begin{figure}[b]
  \begin{center}
    \includegraphics[width=0.48\textwidth,height=0.3\textheight]{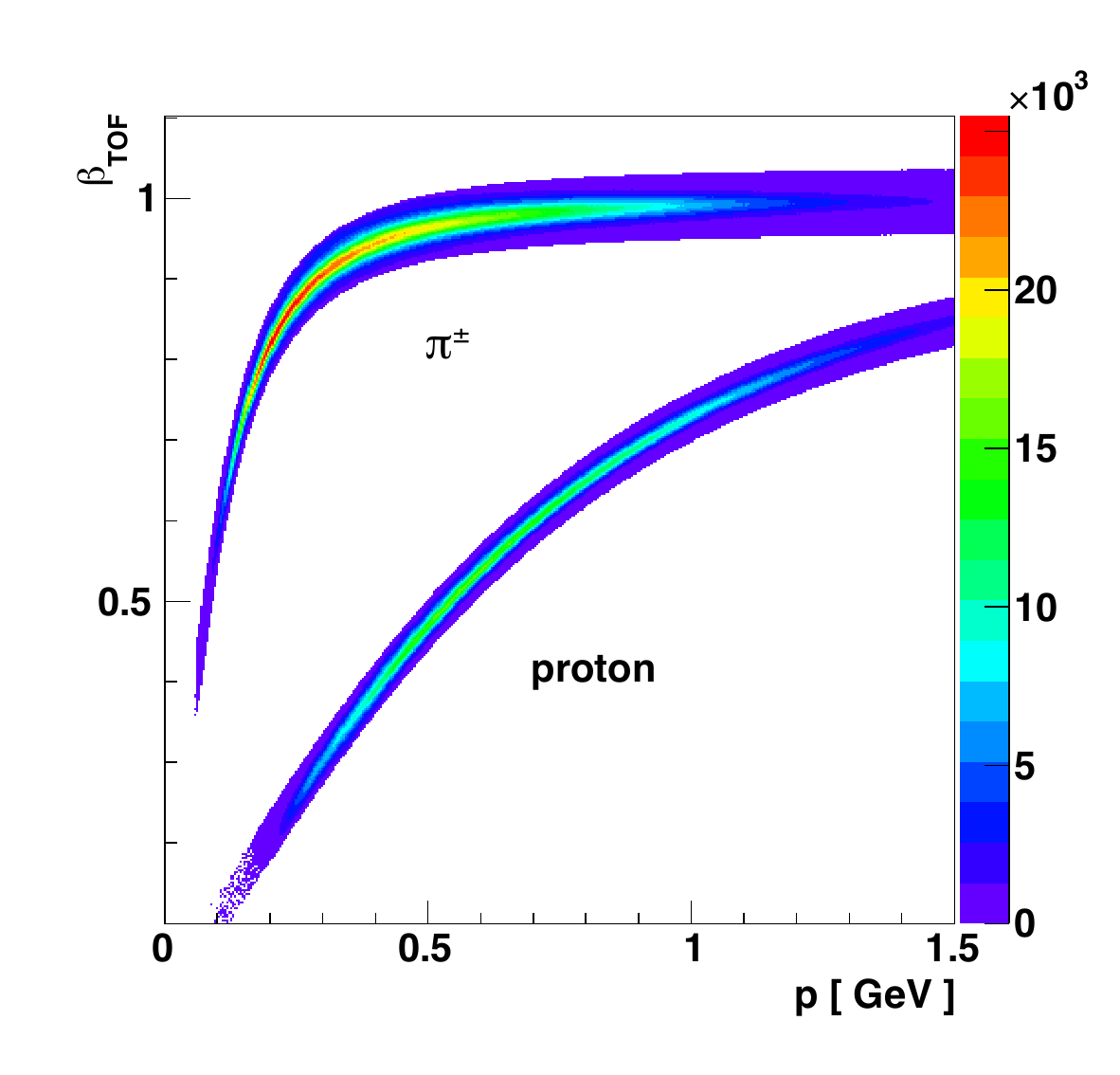}
    \caption{Typical example of a $\beta_{\rm \,TOF}$ versus particle momentum distribution after the 
      $3\sigma$~cuts on $\Delta\beta$.}\label{Figure:DeltaBeta}
  \end{center}
\end{figure}

\begin{figure*}[t]
  \begin{center}
    \includegraphics[width=1.0\textwidth]{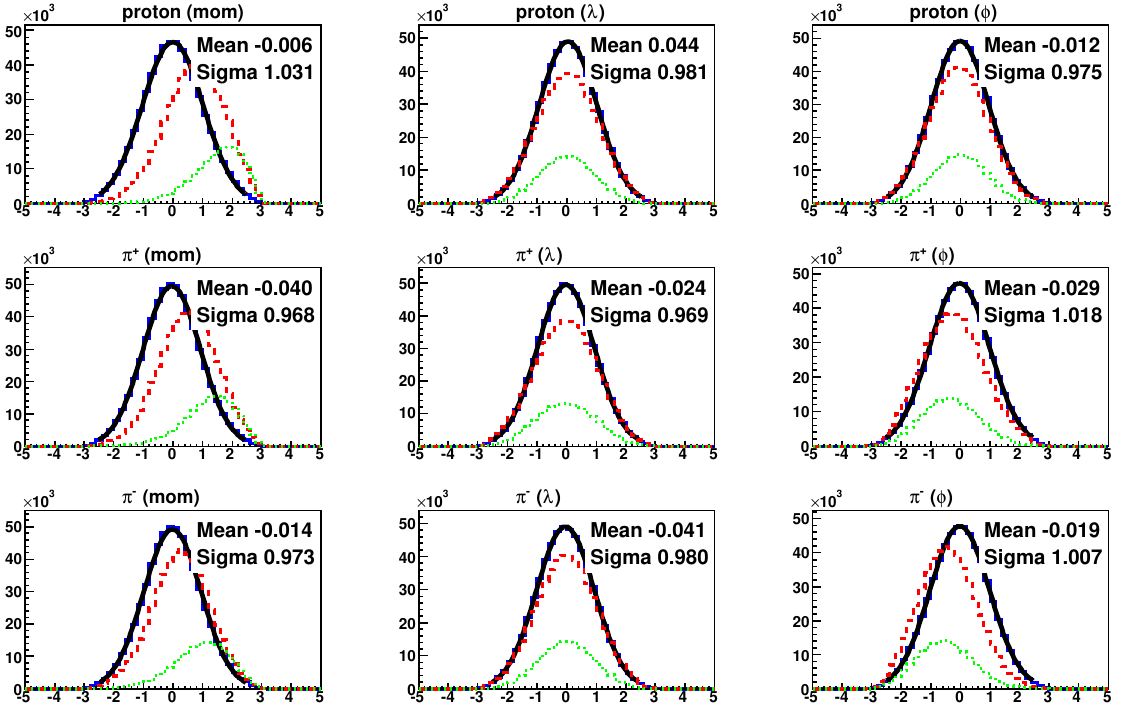}
    \caption{Examples of g9b pull and confidence-level distributions (1.3~GeV coherent edge) from the butanol target
      at various stages in the analysis. The \gr{green}-dotted line was made from the raw data without applying any
      corrections. After the CLAS ELoss package was applied, the \re{red}-dashed line was obtained. A significant
      improvement was observed, in particular for the momentum pulls. Finally, momentum and angle corrections
      were applied and the \bl{blue}-solid histograms were obtained (see text for more details on these corrections).
      These pull and the confidence-level distributions are based on Topology~4, fully exclusive
      $\gamma p\to p\pi^+\pi^-$, with a $5\,\%$ confidence-level cut applied.}\label{Figure:CL_Top05}
  \end{center}
\end{figure*}

Prior to kinematic fitting, the following cuts and event corrections were applied. The incident photons arrived at the
butanol target in 2~ns bunches and, therefore, photon selection cuts were required to identify the correct initial-state
photon. To select the correct photon out of several potential candidates, a cut of $\pm 1$~ns on the coincidence time
(time difference between the event vertex time and the time the photon arrived at the vertex) was applied. This reduced
the initial number of photons from approximately five candidate photons per event to only about 8--10\,\% of all events 
having more than one candidate photon. These events were then discarded. To further minimize the ambiguity in identifying
the correct photon, only those events were considered in which the vertex-timing cut identified the same photon for all
tracks.

For the final-state particle identification, the $\beta$~value of each track was determined from two separate sources:
(1) $\beta_{\rm \,DC} = p/\sqrt{p^2+m^2}$ was measured using the momentum information from the drift chambers and
the PDG mass~\cite{ParticleDataGroup:2022pth} for the particle, and (2) $\beta_{\rm \,TOF}=\frac{v}{c}$ was based on the
velocity information from the time-of-flight (TOF) system~\cite{Smith:1999ii, Mecking:2003zu}. Events were selected
based on $\Delta\beta\,=\,|\beta_{\rm \,DC}\,-\,\beta_{\rm \,TOF}|\,\leq\,3\,\sigma$, where $\sigma$ was the width of
the Gaussian $\Delta\beta$~distributions, which were centered at zero for pions and protons. In this experiment,
$\sigma$~values of 0.013 and 0.01 were measured for the $\pi^\pm$ and proton $\Delta\beta$~distributions, respectively.
This led to a significant improvement in the identification of good final-state tracks and clear bands for protons and
charged pions could be identified in the $\beta_{\,TOF}$ versus momentum distributions (see Fig.~\ref{Figure:DeltaBeta}).
In addition, vertex cuts of $x^2 + y^2 < 9$~cm$^2$ and $-3.0 < z < 3.0$~cm were applied to select events originating
from the butanol target.

\begin{figure}[t]
  \begin{center}
    \includegraphics[width=0.49\textwidth]{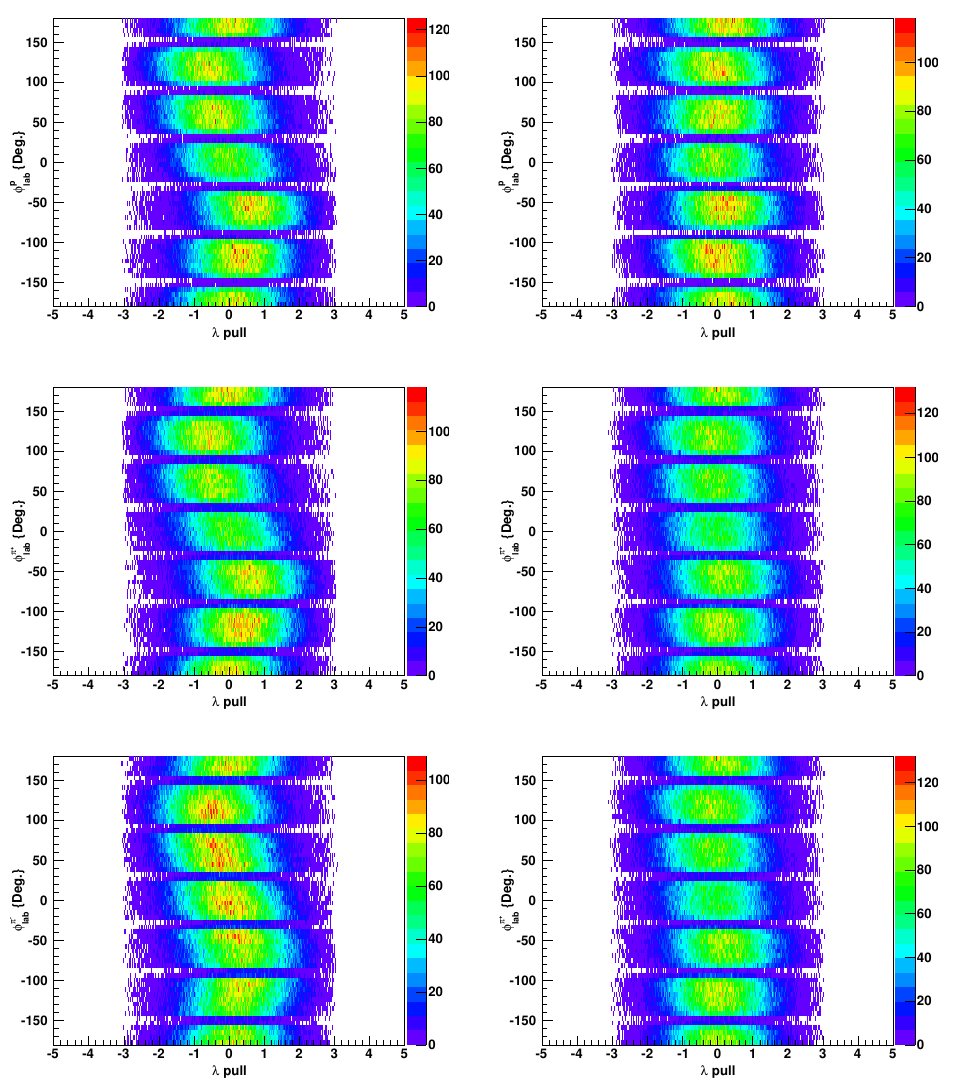}
    \caption{Left column: Mean values of the $\lambda$~pull distributions versus the lab azimuthal angle from
      Topology~4 of FROST-g9b (1.5~GeV coherent-edge data set) for the proton (top), $\pi^{+}$~(center) and
      $\pi^{-}$~(bottom) after applying energy-loss and momentum corrections. The modulations observed in these
      plots indicate that the azimuthal angles (from the momentum vectors) in the lab system require separate corrections
      for each particle. Right column: The same distributions after applying the angle corrections. The initial modulations
      were successfully corrected.}\label{Figure:Modulations}
  \end{center}
\end{figure}

The four-vectors of the selected charged final-state particles were corrected for the energy loss due to the interaction
with materials while traveling through the CLAS volume following the standard CLAS procedure. Small momentum
corrections of a few MeV were also required to correct for factors such as imperfect knowledge of the magnetic field of
the torus magnet and/or misalignments of the drift chambers. The corrections of the $\pi^+$ and proton four-vectors 
were initially determined such that the mass distributions of $X$ in $\gamma p\to p\,X$ and $\gamma p\to p\,\pi^+\,X$
did not have any azimuthal dependence. By using kinematic fitting, these corrections were further fine-tuned and
momentum-dependent corrections for the $\pi^-$ were also found. Figure~\ref{Figure:CL_Top05} shows examples of
CLAS-g9b pull distributions at various stages in the analysis. Particularly, the momentum pulls show big improvements
after applying both energy-loss corrections and additional momentum corrections. All final distributions could be well
described with a Gaussian lineshape. The corresponding mean and width values are given in the figure.

The magnetic holding field surrounding the butanol target was only 0.5~T in strength and, thus, considerably weaker
than the polarizing field of about 5~T. However, the holding field was observed to exert a significant magnetic force on
the charged particles leaving the target by causing $\phi_{\rm \,lab}$\,-\,dependent modulations in the two tracking
parameters $\lambda$ and $\phi$. A $180^\circ$ phase shift was observed for the modulations when the direction of
the holding field was reversed indicating that the modulations did not originate from the beam and target polarizations,
but could be traced back to the holding field. This effect was not taken into account in the initial track reconstruction
and, therefore, led to small deviations from the true trajectories. The correction of these modulations was important
since the extraction of the polarization observables was based on $\phi_{\rm \,lab}$ angular distributions. For this analysis,
$\beta_z$-dependent (= longitudinal momentum dependent) $\phi_{\rm \,lab}$ corrections were developed for the different
particles (proton, $\pi^+$, $\pi^-$) and for the two different signs of the magnetic holding field~\cite{Roy:Dissertation}.
Figure~\ref{Figure:Modulations} shows the observed modulations in the $\lambda$~pull (mean value) versus
$\phi_{\rm \,lab}$~distributions (left column) and the successful removal of the effect after applying the corrections (right
column).    

\subsection*{Event-based signal-background separation\label{ssec:Qfac}}
\begin{figure}[!b]
   \includegraphics[width=0.5\textwidth]{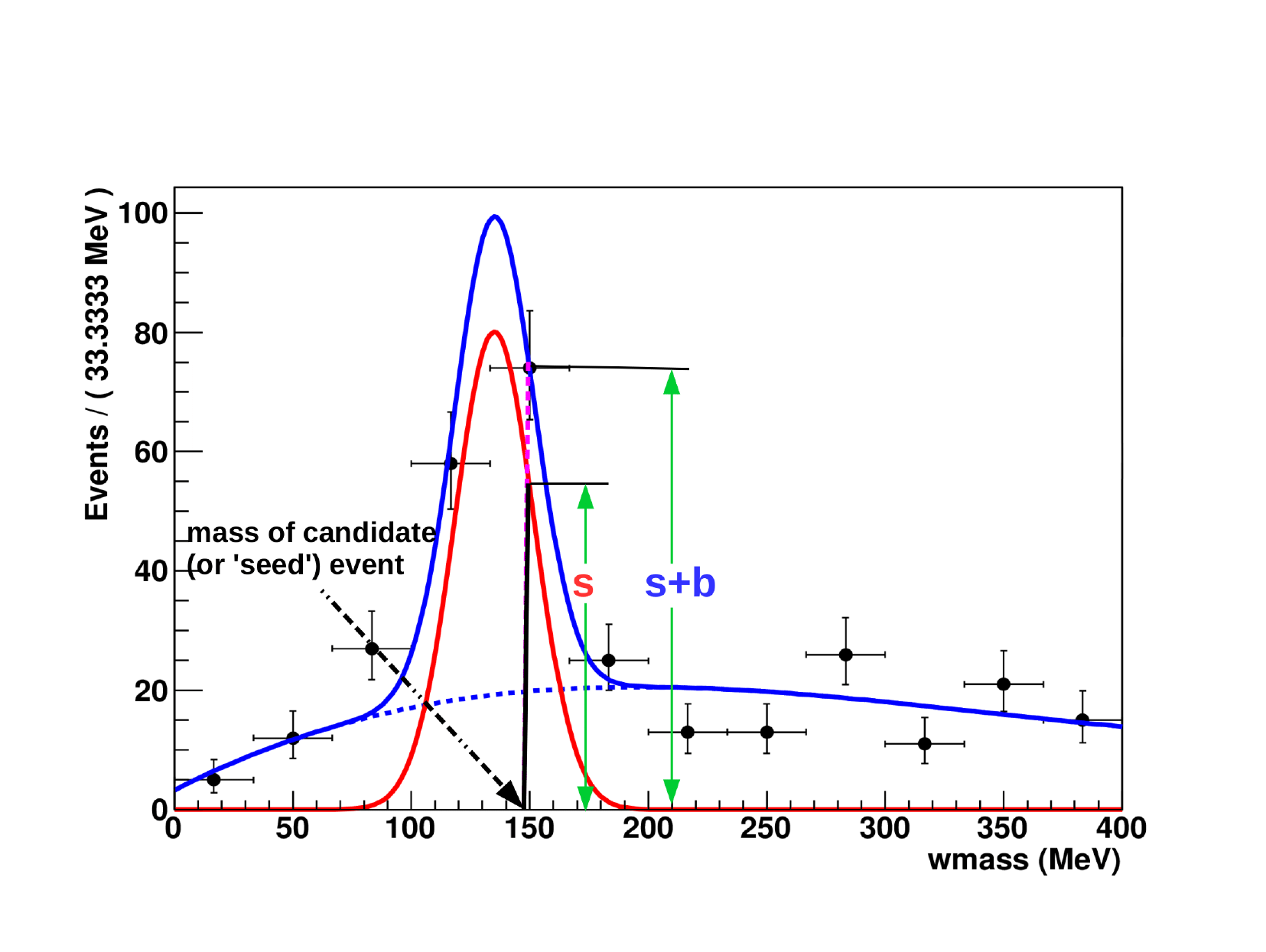}\vspace{-0.5cm}
   \caption{A typical example of a $\pi^-$~missing-mass distribution of the 300~nearest neighbors for an event in the
     energy bin $E_\gamma \in [1050,\,1100]$~MeV. The \bl{blue} solid line represents the total fit, the \re{red} solid line
     the signal (Gaussian pdf), and the \bl{blue} dotted line the background function (second-order Chebychev pdf). The
     $Q$~value of the event was given by $Q=s/(s+b)$, where $s$ was the height of the signal pdf (total pdf) at the
     $\pi^-$~missing-mass of the candidate event and $b$ was the height of the background pdf at the same mass.}
   \label{Figure:Qfac_chi2_fit}
\end{figure}
\begin{figure*}[!ht]
   \includegraphics[width=0.49\textwidth,height=0.29\textheight]{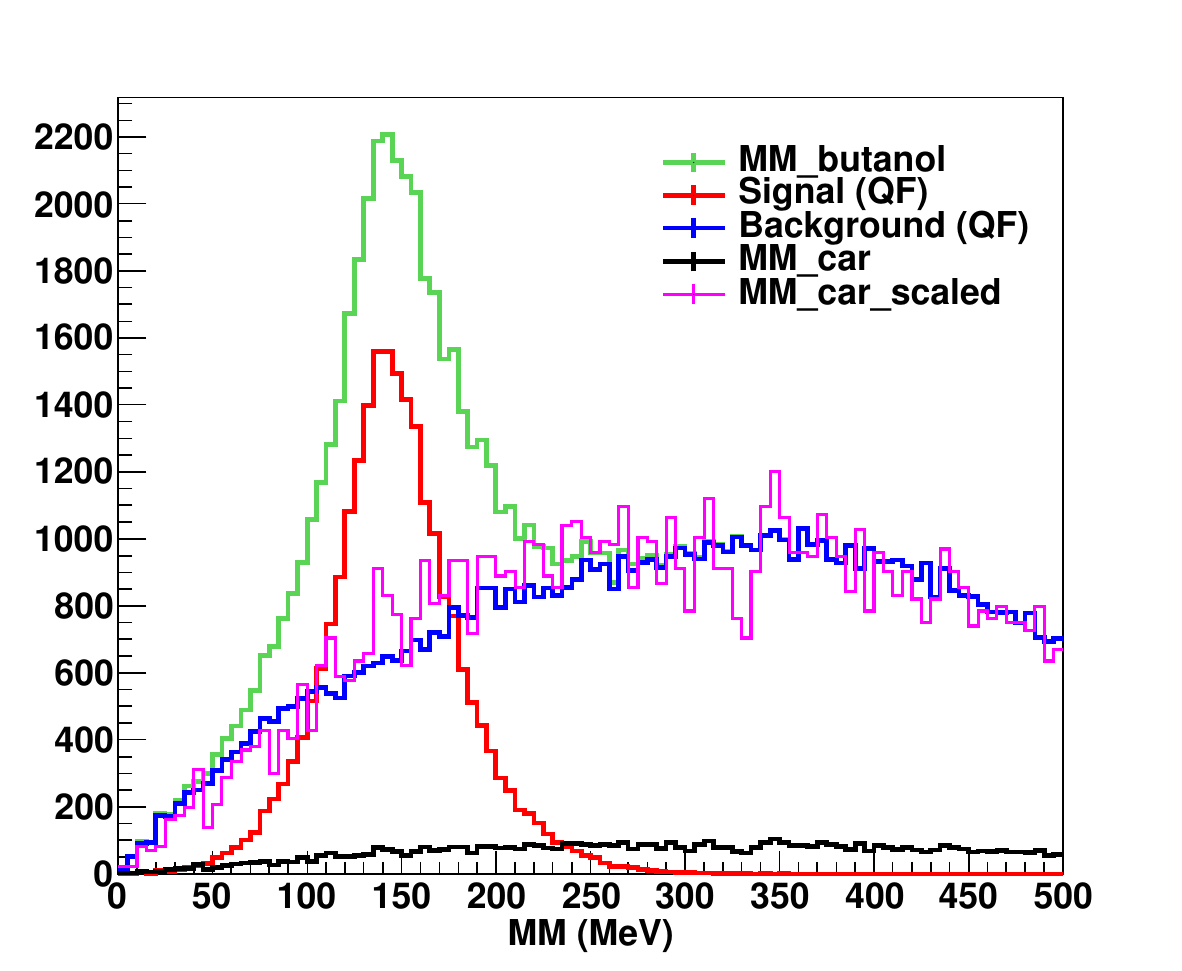}
   \hfill
   \includegraphics[width=0.49\textwidth,height=0.29\textheight]{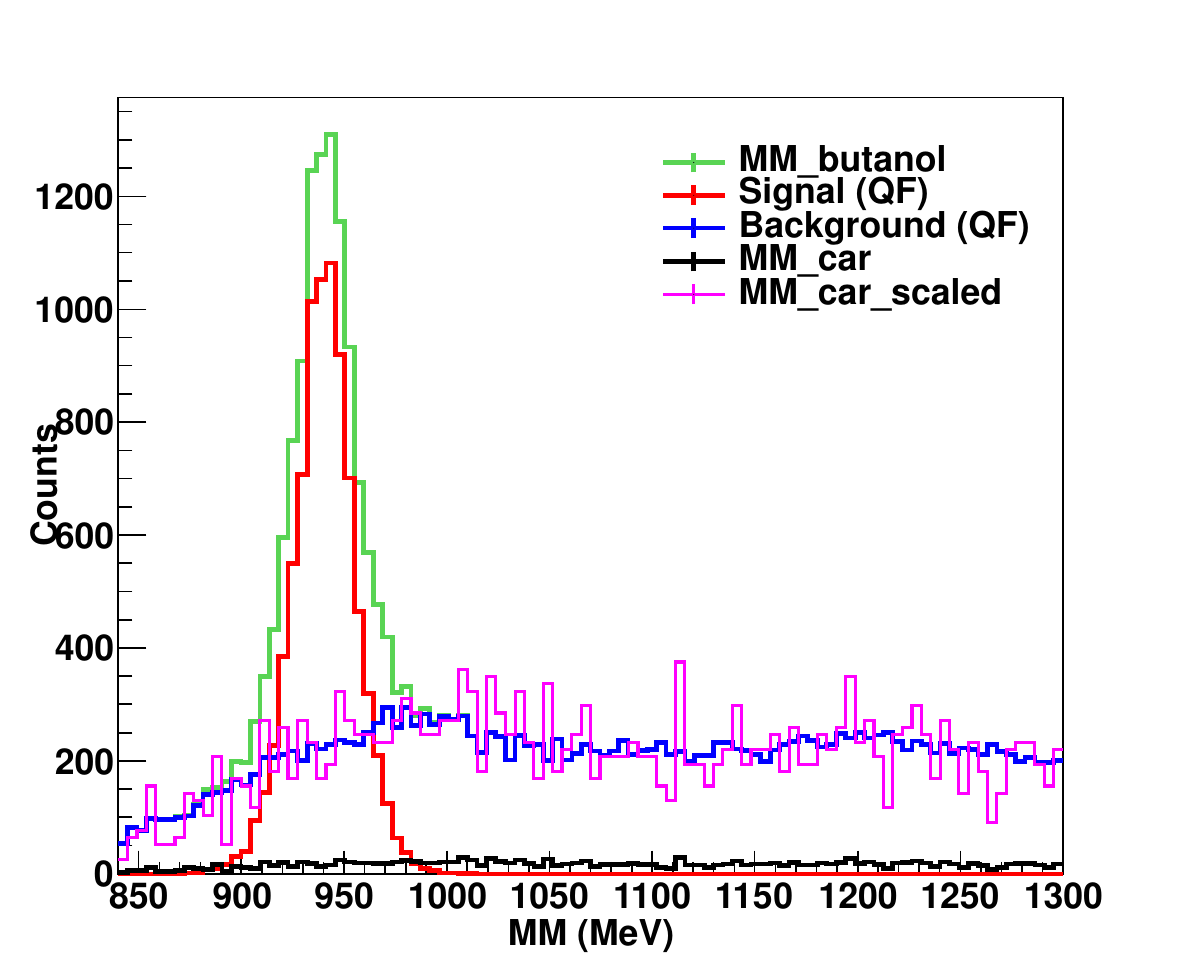}
   \caption{Example of a g9b missing-mass distribution for $E_\gamma \in [1.2,\,1.3]$~GeV. The \gr{green} distribution
     denotes the butanol data, the \re{red} distribution the signal (data weighted with $Q$), and the \bl{blue} distribution
     shows the background (data weighted with $1-Q$). The carbon and scaled carbon distributions are given by the black
     and the \ma{magenta} distributions, respectively. On the left, an example of the missing-$\pi^-$ distribution is
     shown, and on the right, an example of the missing-proton distribution.}\label{Figure:Qfac_mass_histo}
\end{figure*}
The remaining background, consisting mostly of $\pi^+\pi^-$~events originating from bound nucleons of the butanol
target, was separated from signal events using a probabilistic event-based method. This multivariate analysis technique 
is described in detail in Ref.~\cite{Williams:2008sh} and its application in previous CLAS analyses on the measurement of
the $\eta$~photoproduction cross sections and the $\omega$~double-spin asymmetry is detailed in
Refs.~\cite{CLAS:2017yjv,CLAS:2009hpc}. The method determines a weight for each event, denoted as the event $Q$~value,
which denotes the probability for the event being a signal event. These $Q$~values were then used as event weights to
provide any signal distribution, such as angular or mass distributions, and also facilitated the application of event-based
likelihood fits. For this method, the data were divided into data subsets based on their photon energy (binned in $100$-MeV
wide bins) and on their beam and/or target polarization orientations. To determine the $Q$~value for each event in any given
data subset, the 300 kinematically nearest neighbors were selected using a distance metric in the five-dimensional phase
space of the $p\pi^+\pi^-$ final state. In this analysis, the following quantities were chosen: 
\begin{equation*}
{\rm cos}\,\Theta_{\rm \,c.m.}^{\,{\rm p}},\quad {\rm cos}\,\theta_{\,\pi^+}^{\,*},\quad \phi_{\,\pi^+}^{\,*},\quad 
  \phi_{\rm \,lab}^{\,\pi^+}\,,
\end{equation*}
where ${\rm cos}\,\Theta_{\rm \,c.m.}^{\,{\rm p}}$ denotes the cosine of the polar angle of the proton in the center-of-mass
frame, ${\rm cos}\,\theta_{\,\pi^+}^{\,*}$ and $\phi_{\,\pi^+}^{\,*}$ denote the two angles of the $\pi^+$ in the rest frame
of the ($\pi^+\pi^-$) system, and $\phi_{\rm \,lab}^{\,\pi^+}$ is the azimuthal angle of the $\pi^+$ in the lab frame.

Missing-mass distributions were used to separate the signal and background, where the location of the missing mass
peak depended on the topology of the data used. In Topology~1, the missing~$\pi^-$ was fitted and in Topology~2, the
missing~$\pi^+$ was fitted. Topology~3 was not used because the missing proton was indistinguishable from a missing
neutron. In Topology~4, at low energies, the proton was artificially left out because the missing-proton peak was narrower
than the missing-$\pi$ peak. At higher energies above 1600~MeV, both the missing-$\pi$ or missing-proton approach
worked reasonably well, and the artificially missing $\pi^-$ was chosen for this energy range.

This method guaranteed the selection of the 300~nearest neighbors in a very small kinematic region of the
multi-dimensional phase space around the candidate event. Therefore, it was assumed that the signal and background
distributions did not vary rapidly in the selected region and that the missing-mass distribution of these 300~events
determined the $Q$~value of the event. Due to the small sample size of the selected nearest neighbors, an event-based
unbinned maximum likelihood method was implemented to fit the mass distributions. The fit function was defined as:
\begin{equation}
  f(x) \,=\, N\, [ f_{s}\,S(x) \,+\, ( 1 \,-\, f_{s} )\, B(x) ]\,,
  \label{equ:total_function}
\end{equation}
where $S(x)$ denotes the signal and $B(x)$ the background probability density function (pdf). $N$ is a normalization
constant and $f_{s}$ is the signal fraction with a value between 0 and 1. The $Q$~value itself is then given by:
\begin{equation}
  Q \,=\, \frac{s(x)}{s(x) \,+\, b(x)}\,,
  \label{equ:Q_factor}
\end{equation}
where $x$ is the missing-mass of the candidate event, which depends on the topology of the selected event. Moreover,
$s(x) = f_{s} \cdot S(x)$ and $b(x) = (1-f_{s}) \cdot B(x)$. 

A Gaussian function was chosen to describe the signal pdf due to the broad nature of the peak. A second-order Chebychev
polynomial was selected to describe the background pdf. Since the unbinned maximum-likelihood fitting technique did
not provide any goodness-of-fit measure to check the fit quality, the parameter output of each likelihood fit was used
to determine the reduced-$\chi^2$ in a least-squares fit of the missing-mass distribution of the same 300~events. An
example of such a least-squares fit is shown in Figure~\ref{Figure:Qfac_chi2_fit}. The figure also demonstrates how the
$Q$~value was calculated for a candidate event. The choice of a Gaussian for the signal pdf and a second-order Chebychev 
for the background pdf gave the overall best distribution for the reduced-$\chi^2$.

Figure~\ref{Figure:Qfac_mass_histo} shows two examples of missing-mass distributions for all linearly polarized events
in the energy bin $E_{\gamma} \in [1.2,\,1.3]$~GeV, summed over all angles and polarization states. The figure demonstrates
the quality of the applied background-subtraction procedure: the total-mass distribution (green line) was nicely separated
into a Gaussian mass distribution for the signal (red line), obtained by weighting each event with~$Q$, and a smooth
polynomial background (blue line), obtained by weighting each event with ($1-Q$). 

\section{Data Analysis\label{Section:Observables}}
The total cross section, $\sigma$, for $\pi^+\pi^-$~photoproduction using a transversely polarized target can be
expressed in terms of the unpolarized cross section, $\sigma_0$, and a number of polarization observables as:
\begin{equation}
  \begin{split}
    \sigma\, & \, = \,\sigma_0\,\{\,(\, 1\, +\, \bar{\Lambda}_{\rm  t}\,{\rm cos}\,\alpha\,{\bf P_x}\, +\,
    \bar{\Lambda}_{\rm t}\,{\rm sin}\,\alpha\,{\bf P_y}\,)\\[0.5ex]
                   & +\, \bar{\delta}_{\,l}\, [\, {\rm sin}\,2\beta\, (\, {\bf I^{\,s}}\, +\,
    \bar{\Lambda}_{\rm t}\,{\rm cos}\,\alpha\,{\bf P_{\,x}^{\,s}}\, +\, \bar{\Lambda}_{\rm t}\,{\rm sin}\,\alpha\,
       {\bf P_{\,y}^{\,s}})\\[0.5ex]
                   &+\, {\rm cos}\, 2\beta\, (\, {\bf I^{\,c}}\, +\, \bar{\Lambda}_{\rm t}\,{\rm cos}\,\alpha\,{\bf P_{\,x}^{\,c}}\, +
    \bar{\Lambda}_{\rm t}\,{\rm sin}\,\alpha\,{\bf P_{\,y}^{\,c}}\,)\,]\,\}\,,
  \end{split}
  \label{equ:reaction_rate_twopion}
\end{equation}
where the superscripts "s" and "c" refer to sin\,$2\beta$ and cos\,$2\beta$, respectively, $\bar{\delta}_l$ denotes the
average degree of linear-beam polarization (which was observed to be the same for `$+$' and `$-$' target polarizations),
$\bar{\Lambda}_{\rm t}$ denotes the average target polarization (which was also observed to be the same for `$\parallel$'
and `$\perp$' beam polarizations), and the azimuthal angle $\beta$~($\alpha$) is defined as the angle between the photon
beam (target) polarization and the $\hat{x}_{\rm \,event}$-axis in the event frame, as shown in 
Fig.~\ref{Figure:axes_and_angles_def}. Mathematically, 
\begin{equation}
 \beta = \phi= \phi_{\rm \,lab}^{\,\pi^+} - \pi - \phi_0,\quad\,\alpha = \pi - \phi_{\rm \,lab}^{\,\pi^+} + \phi_{\rm \,offset}\,,
 \label{equ:angles_phi_alpha}
\end{equation}
which is also evident from the figure. Here, $\phi_{\rm \,lab}^{\,\pi^+}$ denotes the lab azimuthal angle of the $\pi^+$ and
$\phi_0$~($\phi_{\rm \,offset}$) refers to the orientation of the photon-beam (transversely-polarized target) polarization
with respect to the $\hat{x}_{\rm \,lab}$-axis in the laboratory frame. The definition of the angles and the polarization
observables is analogous to the corresponding definition for the photoproduction of a single-pseudoscalar meson. When
the beam polarization was set to $\parallel$ (or~$\perp$), then $\phi_0 = 0$ (or $\pi/2$)~rad. Similarly, 
$\phi_{\rm \,offset} = 2.025$ (or ($2.025 - \pi$))~rad when the target polarization was set to `$+$'~(or~`$-$'). These
values in radians correspond to $\phi_{\rm \,offset} = 116.1^\circ$ and $-63.9^\circ$, respectively, as discussed in
Section~\ref{Section:ExperimentalSetup}.

The total number of experimentally observed events is related to $\sigma$ as:
\begin{equation}
  N_{\rm \,data}\,=\,\Phi\,C\,\sigma\,,
  \label{equ:Counts}
\end{equation}
where $\Phi$ is the incident photon flux and $C$ denotes the target cross-sectional area.

For the extraction of asymmetries, the absolute value of the photon flux was not required. Rather, the ratios of fluxes
between data sets with different beam-target polarizations were needed to effectively {\it unpolarize} the target in order
to extract the polarization observables, ${\bf I^{\,s}}$ and ${\bf I^{\,c}}$. The flux ratios were determined by using the
information on the total number of reconstructed events from the polyethylene target, which was directly proportional
to the photon flux. This target was chosen since the effects of the magnetic holding field were negligible at the target
location. Events were also counted irrespective of topology so that the ratios were independent of the physics dynamics
involved in the reaction specific to this analysis.
  
\subsection{Extraction of the beam-polarization observables, ${\bf I^{\,s}}$ and ${\bf I^{\,c}}$
  \label{ssec:Likelihood}}  
\begin{figure}[!t]
  \begin{center}
    \includegraphics[width=0.45\textwidth]{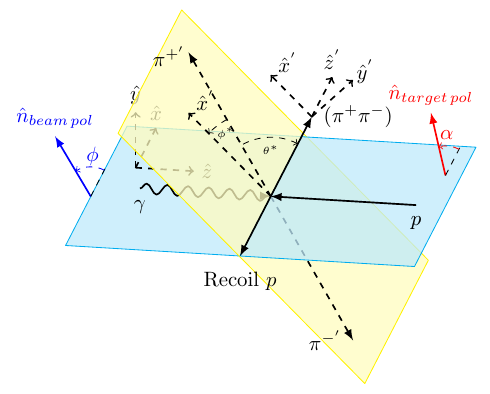}
    \caption{A diagram describing the kinematics of the reaction $\gamma p \to p\,\pi^+\pi^-$. The blue plane
      represents the center-of-mass production plane defined by the initial photon and the recoil proton, whereas the
      yellow plane denotes the decay plane in the $(\pi^+\pi^-)$~rest frame. Moreover, $\phi\,^\ast$ and $\theta\,^\ast$
      describe the azimuthal and polar angle of the $\pi^+$ in the $(\pi^+\pi^-)$~rest frame. Also shown are the beam
      and target polarization orientations and the corresponding azimuthal angles, $\phi$ and $\alpha$ (also in the
      center-of-mass frame).}\label{Figure:KinematicPic_omega}
  \end{center}
\end{figure}

Four independent kinematic variables were required to completely describe the event kinematics in
$\pi^+\pi^-$~photoproduction, as shown in Fig.~\ref{Figure:KinematicPic_omega}. The following variables were chosen: 
the photon energy ($E_\gamma$), the polar and azimuthal angles of the $\pi^+$ in the rest frame ($\theta_{\,\pi^+}$ and
$\phi_{\,\pi^+}$), and the azimuthal angle of the $\pi^+$ ($\phi_{\rm \,lab}^{\,\pi^+}$) in the laboratory frame (not shown
in the figure). The observed modulation in the $\phi_{\rm \,lab}^{\,\pi^+}$~distribution was then used to extract the beam
asymmetry at various ($E_\gamma$, $\theta_{\,\pi^+}$, $\phi_{\,\pi^+}$) bins. An event-based maximum-likelihood fitting
technique was implemented to fit the angular modulations to extract ${\bf I^{\,s}}$ and ${\bf I^{\,c}}$. This technique is
considered more powerful than the conventional binning technique when the data suffer from low statistics since it uses
information from every event, thereby preventing any loss of information due to binning. The method was based on the
principles outlined in Ref.~\cite{CLAS:2016wrl}, which showed its application in a previous CLAS measurement. In 
this analysis, the likelihood (or the joint probability density) of obtaining the experimentally observed 
$\phi_{\rm \,lab}^{\,p}$~angular distribution was expressed in terms of ${\bf I^{\,s}}$ and ${\bf I^{\,c}}$ as the only fit
parameters (see Eq.~\ref{equ:A_sigma_1}-\ref{equ:weights}). To extract ${\bf I^{\,s}}$ and ${\bf I^{\,c}}$ from the FROST
data, the target polarization had to be removed (as detailed below). Maximizing the likelihood function then gave the 
most likely value for~${\bf I^{\,s}}$ and ${\bf I^{\,c}}$.

To nullify the effect of the target polarization to measure ${\bf I^{\,s}}$ and ${\bf I^{\,c}}$, event samples with opposite
target polarization but the same beam polarization were combined using appropriate scale (or normalization)  factors.
The number of $\parallel$~events, $N_\parallel$, after combining data sets with $\parallel$~beam polarization and
different target polarizations (`$+$' or `$-$'), was given by:
\begin{equation}
  N_{\parallel} \,=\, N_{\parallel}^+\,+\, N_1\, N_{\parallel}^-\,,
  \label{equ:N_para_1}
\end{equation}
where $N_1$ was a normalization factor that depended on the photon flux, $\Phi_{\parallel}^+$ and $\Phi_{\parallel}^-$,
and the average degrees of target polarization, $\bar{\Lambda}_{\rm \,t}^+$ and $\bar{\Lambda}_{\rm \,t}^-$, of the two
data sets:
\begin{equation}
  N_1 \,=\, \frac{\Phi_{\parallel}^+\,\bar{\Lambda}_{\rm \,t}^+} {\Phi_{\parallel}^-\,\bar{\Lambda}_{\rm \,t}^-}\,.
\label{equ:norm1}
\end{equation}
Substituting Eq.~\ref{equ:reaction_rate_twopion} and~\ref{equ:Counts} into Eq.~\ref{equ:N_para_1} gives:
\begin{equation}
\begin{split}
  N_{\parallel} \,=\, & \,\,\Phi_\parallel^+ \, C \, \sigma_0\, (1 + \bar{\Lambda}_R)\,\\
     & \times \{ 1 - \bar{\delta}_\parallel \,[\,{\bf I^{\,c}}\,\rm{cos}\,2\phi^{\,\pi^+}_{\,lab}
        +\,\,{\bf I^{\,s}}\,\rm{sin}\,2\phi^{\,\pi^+}_{\,lab}\,] \}\\[0.5ex]
     \,=\, & \,\,\Phi_\parallel^+\,\sigma_\parallel\,,
\end{split}
\label{equ:N_para_2}
\end{equation}
where $\bar{\Lambda}_R$ was defined as $\bar{\Lambda}_R \,=\,\bar{\Lambda}_{\rm \,t}^+ \,/\,\bar{\Lambda}_{\rm \,t}^-$.

Similarly, the number of $\perp$ events, $N_{\perp}$, after combining data sets with $\perp$~beam polarization and
different target polarizations was given by:
\begin{equation}
\begin{split}
  N_{\perp} \,=\, & \,\,\Phi_\perp^+\, C \,\sigma_0\, (1 + \bar{\Lambda}_R)\,\\
    & \times \{ 1 + \bar{\delta}_\perp \, [\,{\bf I^{\,c}}\,\rm{cos}\,2\phi^{\,\pi^+}_{\,lab}
    +\,\,{\bf I^{\,s}}\,\rm{sin}\,2\phi^{\,\pi^+}_{\,lab}\,] \}\\[0.5ex]
    \,=\, & \,\,\Phi_\perp^+\,\sigma_\perp\,.
\end{split}
\label{equ:N_perp_2}
\end{equation}

The asymmetry between $\parallel$~and $\perp$~data can then be expressed as:
\begin{equation}
\begin{split}
   A & \,=\, \frac{N_{\parallel}\,-\,N_{\perp}\,}{N_{\parallel}\,+\,N_{\perp}}
          \,=\, \frac{A^\prime \,+\, \Delta \Phi}{1\,+\,A^\prime\,\Delta \Phi}~,
\end{split}
\label{equ:A_sigma_1}
\end{equation}
where
\begin{align}
  A^\prime &\,=\, \biggl(\frac{\sigma_{\parallel}\,-\,\sigma_{\perp}} {\sigma_{\parallel}\,+\,\sigma_{\perp}}\biggr)
             \,\notag\\
   &\,=\,\frac{\,\tilde{\delta}_l\,(\,{\bf I^{\,s}}\,{\rm sin}\,2\phi^{\,\pi^+}_{\rm \,lab}\,+\,{\bf I^{\,c}}\,{\rm
        cos}\,2\phi^{\,\pi^+}_{\rm \,lab}\,)}{1\,+\,\tilde{\delta}_l\,\Delta\delta_l\,(\,{\bf I^{\,s}}\,{\rm sin}\,
                 2\phi^{\,\pi^+}_{\rm \,lab}\,+\,{\bf I^{\,c}}\,{\rm cos}\,2\phi^{\,\pi^+}_{\rm \,lab}\,)}~,\notag\\[1.5ex]
   \Delta \Phi & ~=~ \frac{\Phi_{\parallel}^+\,-\,\Phi_{\perp}^+} {\Phi_{\parallel}^+\,+\,\Phi_{\perp}^+}~,
   \quad {\rm and}\notag\\[1.5ex]
   \tilde{\delta_l} &\,=\, \frac{ \bar{\delta}_{\parallel}\,+\,\bar{\delta}_{\perp}}{2}\,,\qquad
   \Delta\delta_l \,=\, \frac{\bar{\delta}_{\parallel}\,-\,\bar{\delta}_{\perp}} {\bar{\delta}_{\parallel}\,+\,\bar{\delta}_{\perp}}~.
   \label{equ:A_sigma_2}
\end{align}

The likelihood of obtaining the observed angular distribution in $\phi^{\,\pi^+}_{\rm \,lab}$ in any kinematic bin, using
$A$ from Eqs.~\ref{equ:A_sigma_1}-\ref{equ:A_sigma_2}, was given by:
\begin{equation}
\begin{split}
  -\text{ln}\,L \,=\, -\sum_{i=1}^{N_{\rm{\,total}}}\,w_{\,i} & \,\text{ln}\,(P\,(\text{event}_{\,i})\,)\,,\\[0.5ex]
\end{split}
\end{equation}
\begin{equation*}
\begin{split}
 & {\rm{where}}~P\,(\text{event}_{\,i}) =
\begin{cases}
   ~\frac{1}{2}\,(1\,+\,A) & \text{for $\parallel$ events}\,,\\[1ex]
   ~\frac{1}{2}\,(1\,-\,A) & \text{for $\perp$ events}\,,
\end{cases}
\end{split}
\label{equ:lnL_Sigma}
\end{equation*} 
and $N_{\rm \,total}$ denotes the sum of events over all four beam-target polarization settings used in that kinematic bin.
The weight for each event depended on its $Q_{\rm \,event}$ and the normalization factor for the corresponding data set.
From the above discussion, the weight of the $i^{th}$ event was given by:
\begin{equation}
  w_i \,=\,
  \begin{cases}            
    ~Q_i, & \text{for}~(\parallel,\,+)~{\textrm{or}}~(\perp,\,+)~{\textrm{events}}\,,\\[1.5ex]
    ~Q_i\,\frac{\Phi^{+}_{\parallel}\,\bar{\Lambda}^+_{\rm \,t}}{\Phi^{-}_{\parallel}\,\bar{\Lambda}^-_{\rm \,t}}, 
             & \text{for}~(\parallel,\,-)~{\textrm{events}}\,,\\[2.0ex]
    ~Q_i\,\frac{\Phi^{+}_{\perp}\,\bar{\Lambda}^+_{\rm \,t}}{\Phi^{-}_{\perp}\,\bar{\Lambda}^-_{\rm \,t}}, 
             & \text{for}~(\perp,\,-)~{\textrm{events}}\,.
\end{cases}
\label{equ:weights}
\end{equation}
Minimizing $-$ln\,$L$ yielded the value and the statistical  uncertainty of the polarization observables ${\bf I^{\,s}}$ and
${\bf I^{\,c}}$. This was performed for every ($E_\gamma$, $\theta_{\,\pi^+}$, $\phi_{\,\pi^+}$)~bin. The MINUIT software
package~\cite{cern:minuit} was used for the minimization.

\subsection{Extraction of the target-polarization observables, {\rm $\bf P_{x}$} and {\rm $\bf P_{y}$}
  \label{ssec:TargetAsymmetry}}  
The target-polarization observables {\rm $\bf P_{x}$} and {\rm $\bf P_{y}$} were extracted from the same data using a
transversely-polarized target and an incident linearly polarized photon beam. The same likelihood technique described
in subsection~\ref{ssec:Likelihood} was used to determine this polarization observable. Since the incident photons were
polarized, this beam polarization had to be nullified. 

The number of events with target polarization `$+$', $N^+$, after combining events with different linear polarization
settings, was given by:
\begin{equation}
  N^+ \,=\, N_\parallel^+\,+\, C_\perp^+\, N_\perp^+\,,
\label{equ:N_plus_1}
\end{equation}
where the normalization factor was:
\begin{equation} 
  C^+_{\perp}\,=\,\frac{\Phi_{\parallel}^+}{\Phi_{\perp}^-}\frac{\bar{\delta}_{\parallel}}{\bar{\delta}_{\perp}}.
\label{equ:C_plus_perp}
\end{equation}
By substituting the value of $C^+_\perp$ into Eq.~\ref{equ:N_plus_1} and using Eqs.~\ref{equ:reaction_rate_twopion}
and \ref{equ:Counts}, the number $N^+$ was given by:
\begin{equation}
\begin{split}
  N^+ \,=\, & \,\,\Phi_{\parallel}^+\, \sigma_0\,(1+\bar{\delta}_R)\,\\
   & \{ 1+\bar{\Lambda}_{\rm \,t}^+ \,[\,{\rm \bf P_{x}}\,{\rm cos}\,(\pi\,-\,\phi^{\,\pi^+}_{\rm \,lab}\,+\,\phi_{\rm \,offset})\,\\
   &+\,{\rm \bf P_{y}}\,{\rm sin}\,(\pi\,-\,\phi^{\,\pi^+}_{\rm \,lab}\,+\,\phi_{\rm \,offset})\,]\}\\[0.5ex]
   \,=\, & \,\,\Phi_{\parallel}^+\,\sigma^+\,,
\end{split}
\label{equ:N_plus_2}
\end{equation}
where $\Phi_{\parallel}^+$ was the flux for the data set with target polarization `$+$' and photon beam polarization
`$\parallel$', and $\bar{\delta}_R$ was defined as $\bar{\delta}_R \,=\,\bar{\delta}_{\parallel} \,/\,\bar{\delta}_{\perp}$.

Similarly, the number of events with target polarization `$-$', $N^-$, after combining events with different linear
polarization settings, was given by:
\begin{equation}
\begin{split}
  N^- \,=\, & \,\,\Phi_{\parallel}^-\, \sigma_0\,(1+\bar{\delta}_R)\,\\
    & \{ 1 - \bar{\Lambda}_{\rm \,t}^-
       \,[\,{\rm \bf P_{x}}\,{\rm cos}\,(\pi\,-\,\phi^{\,\pi^+}_{\rm \,lab}\,+\,\phi_{\rm \,offset})\,\\
    &+\,{\rm \bf P_{y}}\,{\rm sin}\,(\pi\,-\,\phi^{\,\pi^+}_{\rm \,lab}\,+\,\phi_{\rm \,offset})\,]\}\\[0.5ex]
    \,=\, & \,\,\Phi_{\parallel}^-\,\sigma^-\,,
\end{split}
\label{equ:N_minus_2}
\end{equation}
where $\Phi_{\parallel}^-$ was the flux for the data set with target polarization `$-$' and photon beam polarization
`$\parallel$'.

The asymmetry between target `$+$'~and `$-$'~data can then be expressed as:
\begin{equation}
  A \,=\, \frac{A^\prime \,+\, \Delta \Phi}{1\,+\,A^\prime\,\Delta \Phi}~,
\label{equ:A_T_1}
\end{equation}
where
\begin{equation*}
\begin{split}
  A^\prime &\,=\, \biggl(\frac{\sigma^+\,-\,\sigma^-} {\sigma^+\,+\,\sigma^-}\biggr)\\[1.5ex]
   & \,=\, \frac{\tilde{\Lambda}\,[\,{\rm \bf P_{x}}\,{\rm cos}\,\alpha\,+\,{\rm \bf
    P_{y}}\,{\rm sin}\,\alpha\,]}{1\,+\,\tilde{\Lambda}\,\Delta\Lambda\,[\,{\rm
    \bf P_{x}}\,{\rm cos}\,\alpha\,+\,{\rm \bf P_{y}}\,{\rm sin}\,\alpha\,]}~,
\end{split}
\end{equation*}
\begin{equation*}
\begin{split}
  \alpha\,&=\,\pi\,-\,\phi^{\,\pi^+}_{\rm \,lab}\,+\,\phi_{\rm \,offset}\,,\\[1.5ex]
  \Delta \Phi \,&=\, \frac{\Phi_{\parallel}^{+}\,-\,\Phi_{\parallel}^-}{\Phi_{\parallel}^{+}\,+\,\Phi_{\parallel}^-}~,\quad
   {\rm and}\\[1.5ex]
   \tilde{\Lambda} &\,=\, \frac{\bar{\Lambda}_{\rm \,t}^+\,+\,\bar{\Lambda}_{\rm \,t}^-}{2}\,,\qquad
   \Delta\Lambda \,=\, \frac{\bar{\Lambda}_{\rm \,t}^+\,-\,\bar{\Lambda}_{\rm \,t}^-}
   {\bar{\Lambda}_{\rm \,t}^+\,+\,\bar{\Lambda}_{\rm \,t}^-}~.
\end{split}
\label{equ:A_T_2}
\end{equation*}

The likelihood of obtaining the observed angular distribution in $\phi^{\,\pi^+}_{\rm \,lab}$ in any kinematic bin,
using~$A$ from Eq.~\ref{equ:A_T_1}, was given by:
\begin{equation}
\begin{split}
  -\text{ln}\,L \,=\, -\sum_{i=1}^{N_{\rm{\,total}}}\,w_{\,i} & \,\text{ln}\,(P\,(\text{event}_{\,i})\,)\,,\\[0.5ex]
\end{split}
\end{equation}
\begin{equation*}
\begin{split}
  & {\rm{where}}~P\,(\text{event}_{\,i})=
\begin{cases}
  ~\frac{1}{2}\,(1\,+\,A) & \text{for `$+$' events}\,,\\[1ex]
  ~\frac{1}{2}\,(1\,-\,A) & \text{for `$-$' events}\,,
\end{cases}
\end{split}
\label{equ:lnL_T}
\end{equation*} 
and $N_{\rm \,total}$ denotes the sum of events over the two target-polarization settings used in that kinematic bin.
The weight of the $i^{th}$ event was:
\begin{equation}
  w_i \,=\,
\begin{cases}
  ~Q_i, & {\rm if}~(\parallel,\,+)~{\rm or}~(\parallel,\,-)~{\rm event}\\[0.5ex]
  ~Q_i\,\frac{\Phi_{\parallel}^+}{\Phi_{\perp}^+}\,\frac{\bar{\delta}_{\parallel}}{\bar{\delta}_{\perp}} &
  {\rm if}~(\perp,\,+)~{\rm event}\\[0.5ex]
  ~Q_i\,\frac{\Phi_{\parallel}^-}{\Phi_{\perp}^-}\,\frac{\bar{\delta}_{\parallel}}{\bar{\delta}_{\perp}} &
  {\rm if}~(\perp,\,-)~{\rm event}
\end{cases}
\end{equation}

\noindent
The observables {\rm $\bf P_{x}$} and {\rm $\bf P_{y}$} were then extracted by minimizing -ln\,$L$.

\subsection{Extraction of the double-polarization observables, {\rm $\bf P_{\,x,\,y}^{\,s,\,c}$}
  \label{ssec:DoubleAsymmetry}}  
The double-polarization observables {\rm $\bf P_{\,x,\,y}^{\,s,\,c}$} were extracted from the same data using a
transversely-polarized target and an incident linear-polarized photon beam. The same likelihood technique described
in subsection IV A was used to determine this polarization observable. 

Combining data sets was done in a way such that the neither the beam polarization nor the target polarization would be
nullified. Combining $N_\parallel^{+}$ with $N_\perp^{-}$ data with appropriate normalization factors gave:
\begin{equation}
    \begin{split}
        N_1\,&=\,\,N_{\parallel}^{+}+C_1\,N_{\perp}^{-}\\[1ex]
        &=\,\,\Phi_{\parallel}^+\sigma_0\,\{ 1 + \bar{\Lambda}_R \\
        &\qquad + \bar{\delta}_{\perp}\,(\bar{\delta}_R-\bar{\Lambda}_R)\,[\,{\rm \bf I^{\,s}}\,{\rm sin}\,2\beta\,+
        {\rm \bf I^{\,c}}\,{\rm cos}\,2\beta\,]\\
        &\qquad + \,\bar{\Lambda}^+_{\rm \,t}\,(\bar{\delta}_{\parallel}+\bar{\delta}_{\perp})\,[\,{\rm sin}\,2\beta\,
        ({\rm\bf P_{\,x}^{\,s}}\,{\rm cos}\,\alpha+{\rm\bf P_{\,y}^{\,s}}\,{\rm sin}\,\alpha)\\
        &\qquad+\,{\rm cos}\,2\beta\,({\rm\bf P_{\,x}^{\,c}}\,{\rm cos}\,\alpha+{\rm\bf P_{\,y}^{\,c}}\,
        {\rm sin}\,\alpha)\,]\}\,,
    \end{split}
    \label{equ:N_1_sigma_complex}
\end{equation}
where
\begin{equation*}
    \begin{split}
        C_1\,&=\,\frac{\Phi_{\parallel}^+}{\Phi_{\perp}^-}\frac{\bar{\Lambda}^+_{\rm \,t}}{\bar{\Lambda}^-_{\rm \,t}}\,,\\[1.5ex]
        \bar{\delta}_R\,&=\,\frac{\bar{\delta}_{\parallel}}{\bar{\delta}_{\perp}}~,\qquad\text{and} \qquad\bar{\Lambda}_R\,
        =\,\frac{\bar{\Lambda}^+_{\rm \,t}}{\bar{\Lambda}^-_{\rm \,t}}~.
    \end{split}
\end{equation*}

\noindent
$\Phi_{\parallel}^+$ was the flux for the data set with target polarization `$+$' and photon beam polarization `$\parallel$'.
The angles $\alpha$ and $\beta$ were defined as $\alpha\,=\,\pi\,-\,\phi^{\,\pi^+}_{\rm \,lab}\,+\,\phi_{\rm \,offset}$ and
$\beta\,=\,\phi^{\,\pi^+}_{\rm \,lab}$.\\[-1.5ex]

In the term $\bar{\delta}_{\perp}\,(\bar{\delta}_R-\bar{\Lambda}_R)\,[\,{\rm \bf I^{\,s}}\,{\rm sin}\,2\beta\,+
{\rm \bf I^{\,c}}\,{\rm cos}\,2\beta\,]$, $\bar{\delta}_R$ and $\bar{\Lambda}_R$ were $\thicksim$ 1, and the remaining
variables were smaller than 1; thus, this term could be discarded. Then $N_1$ was simplified to:
\begin{align}
    N_1\,&=\,\,\Phi_{\parallel}^+\sigma_0\,\{ 1 + \bar{\Lambda}_R \notag\\
    &\qquad+\,\bar{\Lambda}^+_{\rm \,t}\,(\bar{\delta}_{\parallel}+\bar{\delta}_{\perp})\,[\,{\rm sin}\,2\beta\,
      ({\rm\bf P_{\,x}^{\,s}}\,{\rm cos}\,\alpha+{\rm\bf P_{\,y}^{\,s}}\,{\rm sin}\,\alpha)\notag\\
    &\qquad+\,{\rm cos}\,2\beta\,({\rm\bf P_{\,x}^{\,c}}\,{\rm cos}\,\alpha+{\rm\bf P_{\,y}^{\,c}}\,{\rm sin}\,\alpha)\,]\},
      \notag\\
    &=\,\,\Phi_{\parallel}^+\sigma_1\,.
    \end{align}
    \label{equ:N_1_sigma_simp}
Similarly, combining $N_\perp^{+}$ with $N_\parallel^{-}$ gave:
    \begin{align}
        N_2\,&=\,\,N_{\perp}^{+}+C_2\,N_{\parallel}^{-}\notag\\[1ex]
        &=\,\,\Phi_{\perp}^+\sigma_0\,\{ 1 + \bar{\Lambda}_R \notag\\
        &\qquad-\,\bar{\Lambda}^+_{\rm \,t}\,(\bar{\delta}_{\parallel}+\bar{\delta}_{\perp})\,[\,{\rm sin}\,2\beta\,
          ({\rm\bf P_{\,x}^{\,s}}\,{\rm cos}\,\alpha+{\rm\bf P_{\,y}^{\,s}}\,{\rm sin}\,\alpha)\notag\\
        &\qquad+\,{\rm cos}\,2\beta\,({\rm\bf P_{\,x}^{\,c}}\,{\rm cos}\,\alpha+{\rm\bf P_{\,y}^{\,c}}\,
          {\rm sin}\,\alpha)\,]\},\notag\\
        &=\,\,\Phi_{\perp}^+\sigma_2\,,
    \end{align}
    \label{equ:N_2_sigma_simp}
where
\begin{equation*}
        C_2\,=\,\frac{\Phi_{\perp}^+}{\Phi_{\parallel}^-}\frac{\bar{\Lambda}^+_{\rm \,t}}{\bar{\Lambda}^-_{\rm \,t}}~.\\[0.5ex]
\end{equation*}

\noindent
The asymmetry between `1' and `2' data was given by:
\begin{equation}
   A \,=\, \frac{A^\prime \,+\, \Delta \Phi}{1\,+\,A^\prime\,\Delta \Phi}~,
\label{equ:A_double_1}
\end{equation}
where
\begin{equation*}
\begin{split}
  A^\prime &\,=\, \biggl(\frac{\sigma_1\,-\,\sigma_2} {\sigma_1\,+\,\sigma_2}\biggr)\\[1.5ex]
   & \,=\, \frac{\bar{\Lambda}^+_{\rm \,t}\,(\bar{\delta}_{\parallel}+\bar{\delta}_{\perp})}{1+\bar{\Lambda}_R}\,
   (\,{\rm\bf P_{\,x}^{\,s}}\,{\rm sin}\,2\beta\,{\rm cos}\,\alpha+{\rm\bf P_{\,y}^{\,s}}\,{\rm sin}\,2\beta\,{\rm sin}\,
   \alpha\\[1ex]
   &\qquad+{\rm\bf P_{\,x}^{\,c}}\,{\rm cos}\,2\beta\,{\rm cos}\,\alpha+{\rm\bf P_{\,y}^{\,c}}\,{\rm cos}\,2\beta\,
   {\rm sin}\,\alpha),\\[1.5ex]
   \Delta \Phi \,&=\, \frac{\Phi_{\parallel}^{+}\,-\,\Phi_{\perp}^-}{\Phi_{\parallel}^{+}\,+\,\Phi_{\perp}^-}.
\end{split}
\label{double_asymmetry}
\end{equation*}

The likelihood of obtaining the observed angular distribution in $\phi^{\,\pi^+}_{\rm \,lab}$ in any kinematic bin,
using $A$ from Eq.~\ref{equ:A_double_1}, was given by:
\begin{equation}
\begin{split}
  -\text{ln}\,L \,=\, -\sum_{i=1}^{N_{\rm{\,total}}}\,w_{\,i} & \,\text{ln}\,(P\,(\text{event}_{\,i})\,)\,,\\[0.5ex]
\end{split}
\end{equation}
where
\begin{equation*}
\begin{split}
  & P\,(\text{event}_{\,i})=
\begin{cases}
  ~\frac{1}{2}\,(1\,+\,A) & {\rm for}~(\parallel,\,+)~{\rm or}~(\perp,\,-)~{\rm events}\,,\\[1ex]
  ~\frac{1}{2}\,(1\,-\,A) & {\rm for}~(\parallel,\,-)~{\rm or}~(\perp,\,+)~{\rm events}\,,
\end{cases}
\end{split}
\label{equ:lnL_double}
\end{equation*} 
and $N_{\rm \,total}$ denotes the sum of events over the two target-polarization settings used in that kinematic bin.
The weight of the $i^{th}$ event was:
\begin{equation}
w_i \,=\,
\begin{cases}
   ~Q_i & {\rm if}~(\parallel,\,+)~{\rm or}~(\perp,\,+)~{\rm event}\,,\\[0.5ex]
   ~Q_i\,\frac{\Phi_{\parallel}^{+}}{\Phi_{\perp}^{-}}\,\bar{\Lambda}_R & {\rm if}~(\perp,\,-)~{\rm event}\,,\\[0.5ex]
   ~Q_i\,\frac{\Phi_{\perp}^{+}}{\Phi_{\parallel}^{-}}\,\bar{\Lambda}_R & {\rm if}~(\parallel,\,-)~{\rm event}\,.
\end{cases}
\end{equation}

\noindent
The observables {\rm $\bf P_{\,x,\,y}^{\,s,\,c}$} were then  extracted by minimizing -ln\,$L$.

\section{Results\label{Section:Results}}
This section presents the experimental results for the four single-polarization asymmetries ${\bf I^{\,s,c}}$ (beam
asymmetries) as well as the ${\bf P_{\,x,y}}$ (target asymmetries), and the four beam-target double-polarization
observables ${\bf P_{\,x,y}^{\,s,c}}$ in the photoproduction of a $\pi^+\pi^-$~pair off the proton. To provide a better
overview, example distributions are presented for the single-polarization observables and the full set of results for the
double-polarization observables.

\begin{figure*}[t]
  \begin{center}
    \includegraphics[width=1.0\textwidth]{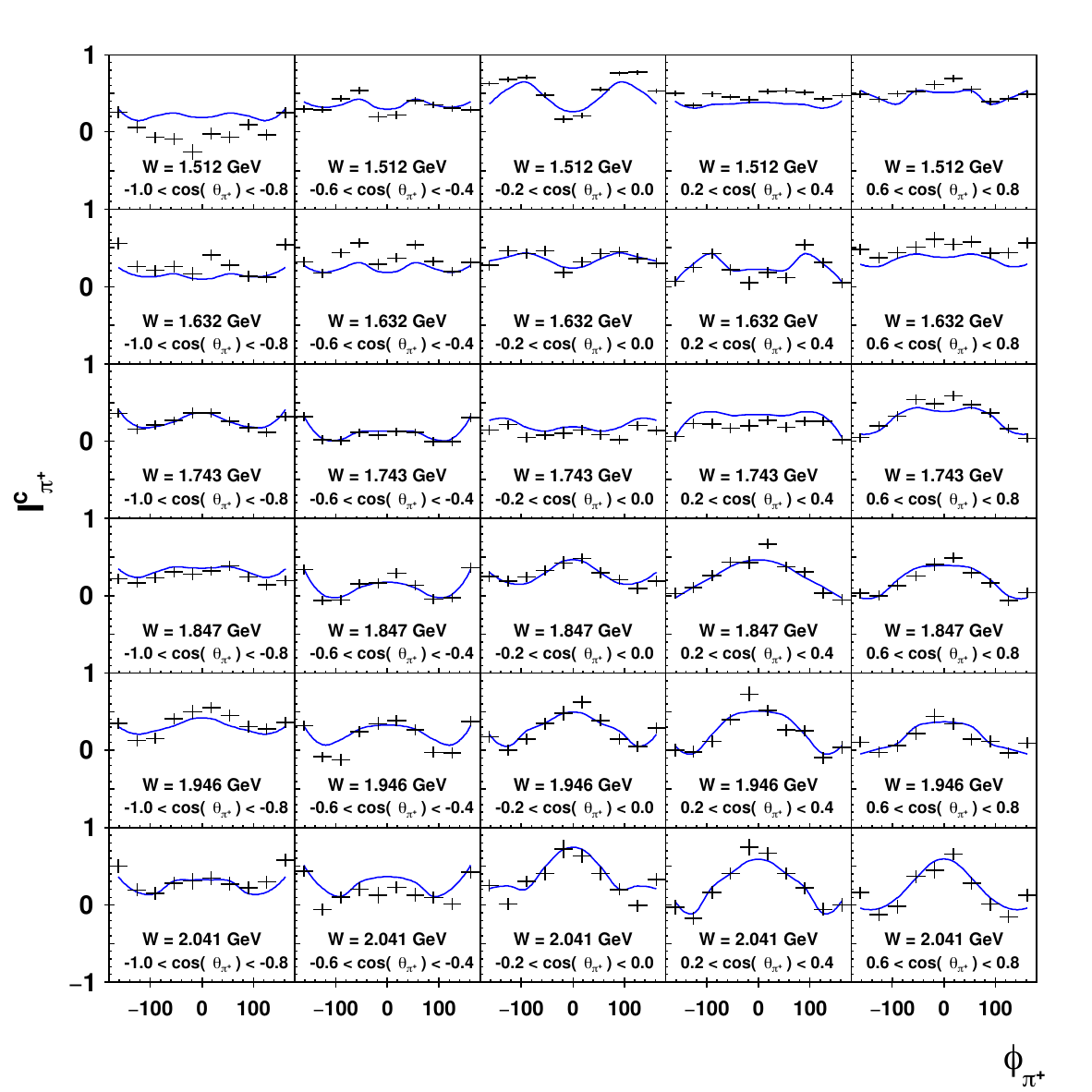}
    \caption{Results for the beam asymmetry, $\rm\bf I^{\,c}$, using a linearly polarized photon beam in the reaction
      $\gamma p\to p\pi^+\pi^-$. The data have been analyzed for 100-MeV-wide energy bins in the incident photon
      energy, $E_\gamma\in [700,1800]$~MeV, and are shown for every other energy bin in the corresponding center-of-mass
      energy~$W$. The data points are presented in 0.2-wide~cos\,$\theta$~bins and are shown versus~$\phi$. These two
      quantities denote the polar and azimuthal angle of the $\pi^+$~meson in the rest frame (helicity frame) of the
      $\pi^+\pi^-$~system; again, every other cos\,$\theta$~bin is shown. The blue solid line denotes the BnGa-PWA
      solution.}\label{Figure:I_c}
  \end{center}
\end{figure*}

\begin{figure*}[t]
  \begin{center}
    \includegraphics[width=1.0\textwidth]{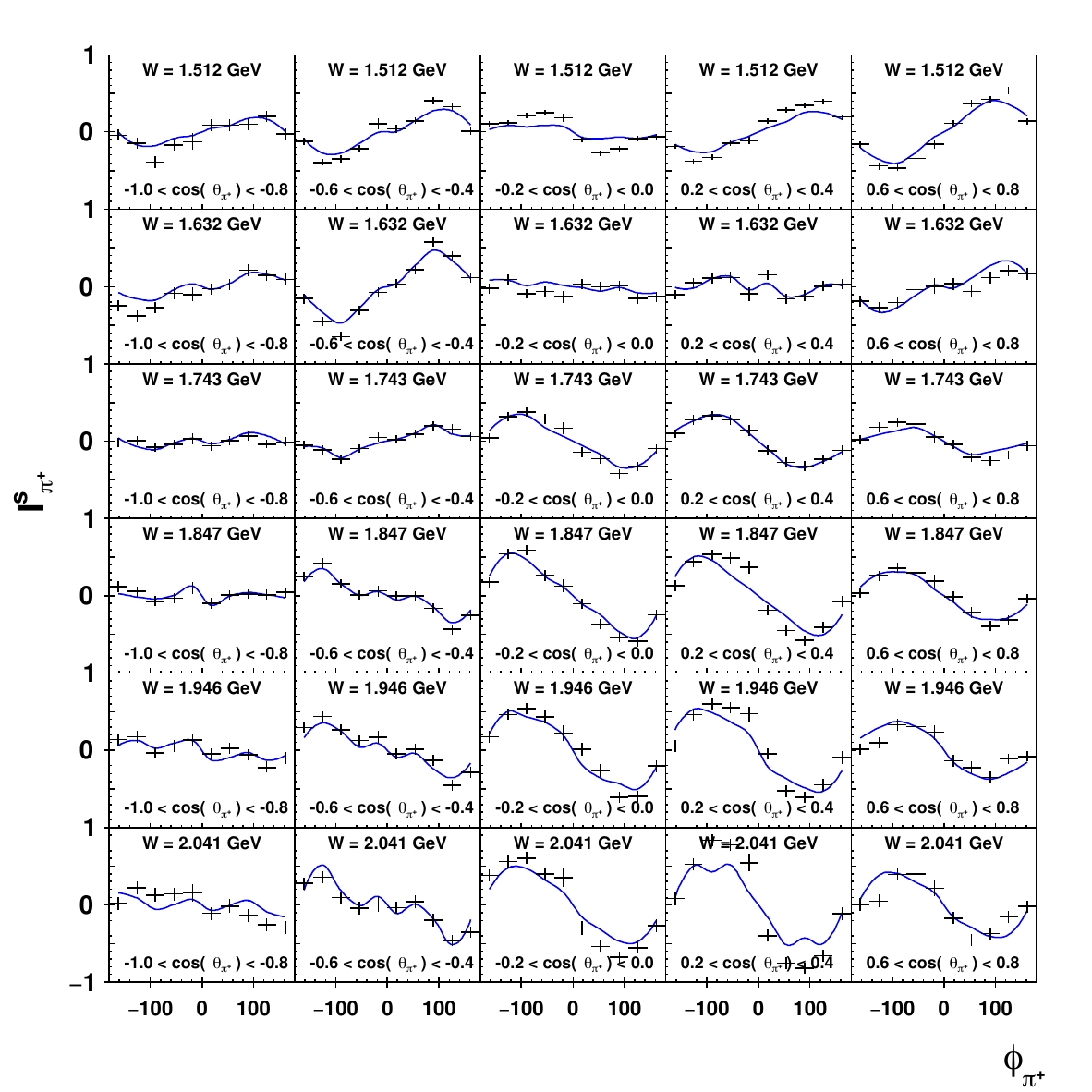}
    \caption{Results for the beam asymmetry, $\rm\bf I^{\,s}$, using a linearly polarized photon beam in the reaction
      $\gamma p\to p\pi^+\pi^-$. The data have been analyzed for 100-MeV-wide energy bins in the incident photon
      energy, $E_\gamma\in [700,1800]$~MeV, and are shown for every other energy bins in the corresponding center-of-mass
      energy~$W$. The data points are presented in 0.2-wide~cos\,$\theta$~bins and are shown versus~$\phi$. These two
      quantities denote the polar and azimuthal angle of the $\pi^+$~meson in the rest frame (helicity frame) of the
      $\pi^+\pi^-$~system; again, every other cos\,$\theta$~bin is shown. The blue solid line denotes the BnGa-PWA
      solution.}\label{Figure:I_s}
  \end{center}
\end{figure*}

\begin{figure*}[t]
  \begin{center}
    \includegraphics[width=1.0\textwidth]{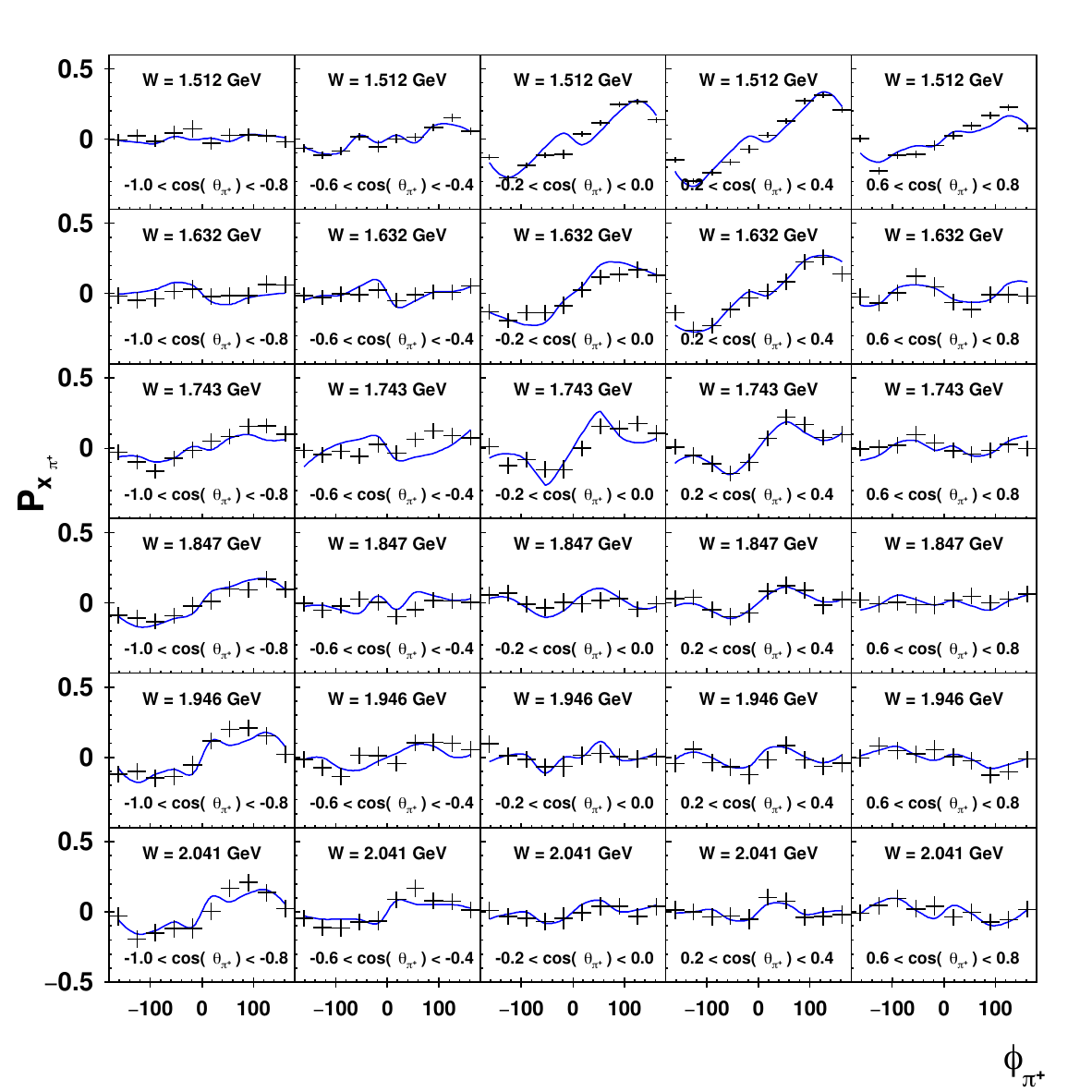}
    \caption{Results for the target asymmetry, $\rm\bf P_{x}$, using a transversely polarized target in the reaction
      $\gamma p\to p\pi^+\pi^-$. The data have been analyzed for 100-MeV-wide energy bins in the incident photon
      energy, $E_\gamma\in [700,1800]$~MeV, and are shown for every other energy bins in the corresponding center-of-mass
      energy~$W$. The data points are presented in 0.2-wide~cos\,$\theta$~bins and are shown versus~$\phi$. These two
      quantities denote the polar and azimuthal angle of the $\pi^+$~meson in the rest frame (helicity frame) of the
      $\pi^+\pi^-$~system; again, every other cos\,$\theta$~bin is shown. The blue solid line denotes the BnGa-PWA
      solution.}\label{Figure:P_x}
  \end{center}
\end{figure*}

\begin{figure*}[t]
  \begin{center}
    \includegraphics[width=1.0\textwidth]{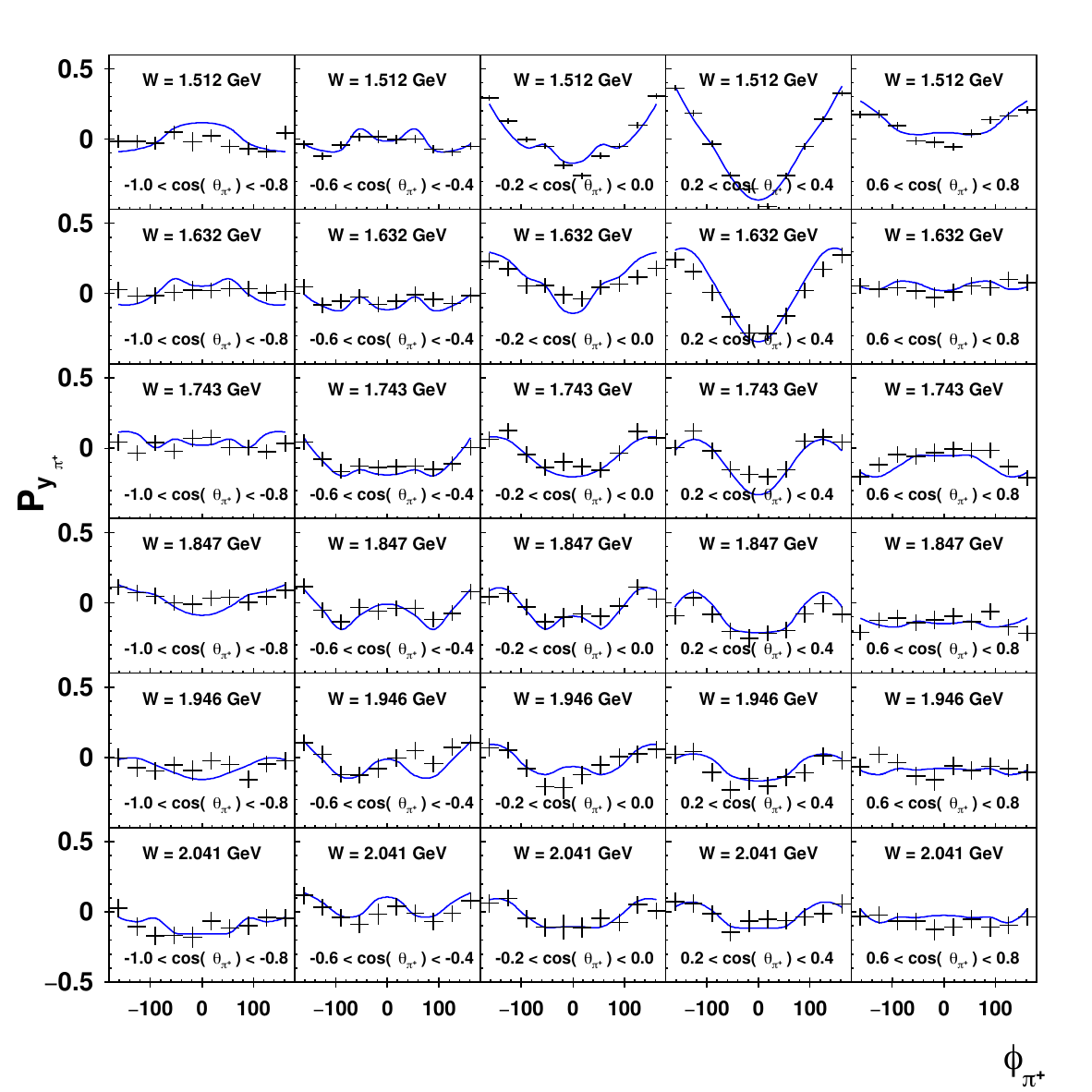}
    \caption{Results for the target asymmetry, $\rm\bf P_y$, using a transversely polarized target in the reaction
      $\gamma p\to p\pi^+\pi^-$. The data have been analyzed for 100-MeV-wide energy bins in the incident photon
      energy, $E_\gamma\in [700,1800]$~MeV, and are shown for every other energy bins in the corresponding center-of-mass
      energy~$W$. The data points are presented in 0.2-wide~cos\,$\theta$~bins and are shown versus~$\phi$. These two
      quantities denote the polar and azimuthal angle of the $\pi^+$~meson in the rest frame (helicity frame) of the
      $\pi^+\pi^-$~system; again, every other cos\,$\theta$~bin is shown. The blue solid line denotes the BnGa-PWA
      solution.}\label{Figure:P_y}
  \end{center}
\end{figure*}

\begin{figure*}[t]
  \begin{center}
    \includegraphics[width=1.0\textwidth]{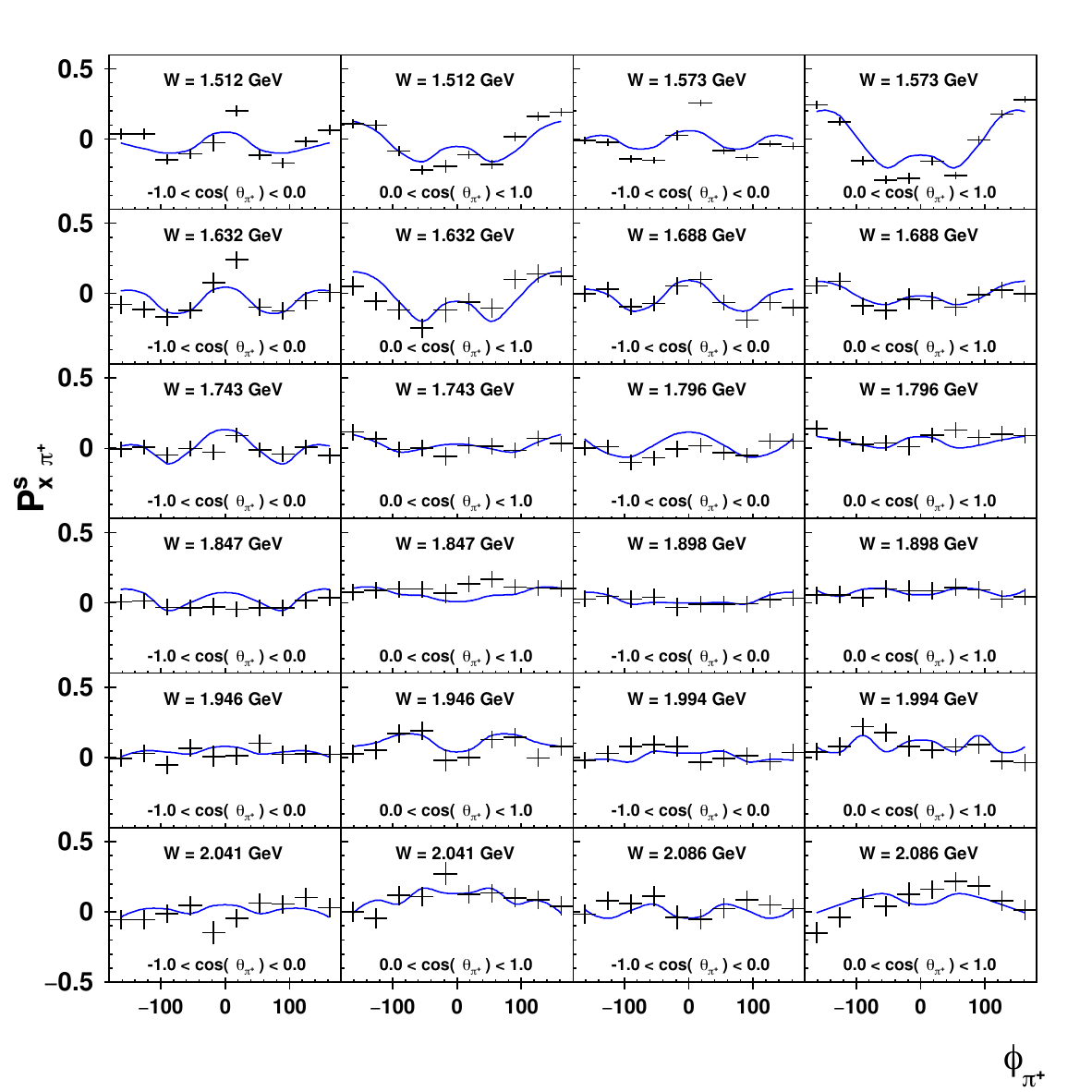}
    \caption{Full results for the beam-target observable, $\rm\bf P_x^s$, using a linearly polarized photon beam and
      a transversely polarized target in the reaction $\gamma p\to p\pi^+\pi^-$. The data have been analyzed for
      100-MeV-wide energy bins in the incident photon energy, $E_\gamma\in [700,1800]$~MeV, and are shown for
      bins in the corresponding center-of-mass energy~$W$. The data points are presented in the backward
      (cos\,$\theta < 0$) and forward direction (cos\,$\theta > 0$) are shown versus~$\phi$. These two quantities
      denote the polar and azimuthal angle of the $\pi^+$~meson in the rest frame (helicity frame) of the
      $\pi^+\pi^-$~system. The blue solid line denotes the BnGa-PWA solution.}\label{Figure:P_x_s}
  \end{center}
\end{figure*}

\begin{figure*}[t]
  \begin{center}
    \includegraphics[width=1.0\textwidth]{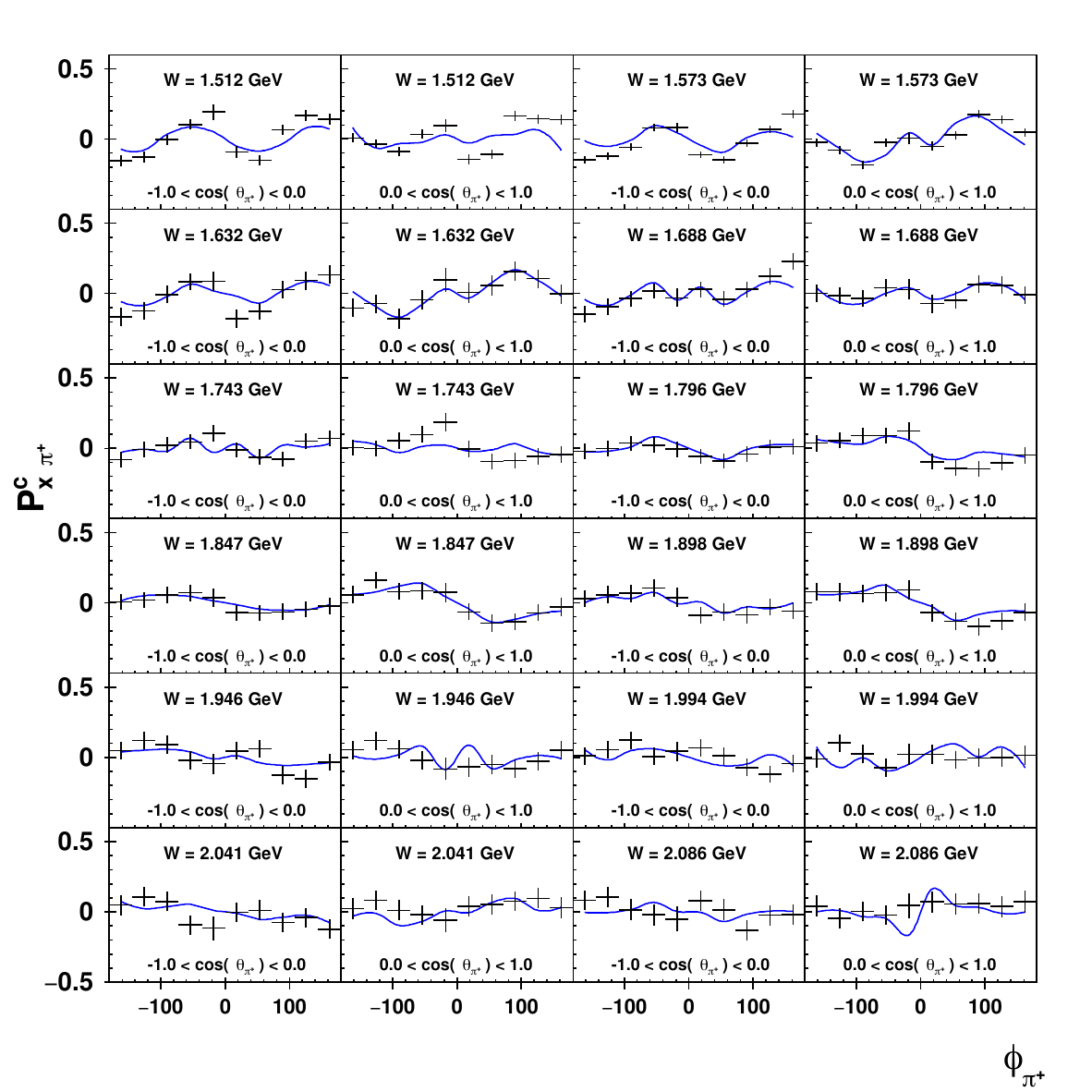}
    \caption{Full results for the beam-target observable, $\rm\bf P_x^c$, using a linearly polarized photon beam and
      a transversely polarized target in the reaction $\gamma p\to p\pi^+\pi^-$. The data have been analyzed for
      100-MeV-wide energy bins in the incident photon energy, $E_\gamma\in [700,1800]$~MeV, and are shown for
      bins in the corresponding center-of-mass energy~$W$. The data points are presented in the backward
      (cos\,$\theta < 0$) and forward direction (cos\,$\theta > 0$) are shown versus~$\phi$. These two quantities
      denote the polar and azimuthal angle of the $\pi^+$~meson in the rest frame (helicity frame) of the
      $\pi^+\pi^-$~system. The blue solid line denotes the BnGa-PWA solution.}\label{Figure:P_x_c}
  \end{center}
\end{figure*}

\begin{figure*}[t]
  \begin{center}
    \includegraphics[width=1.0\textwidth]{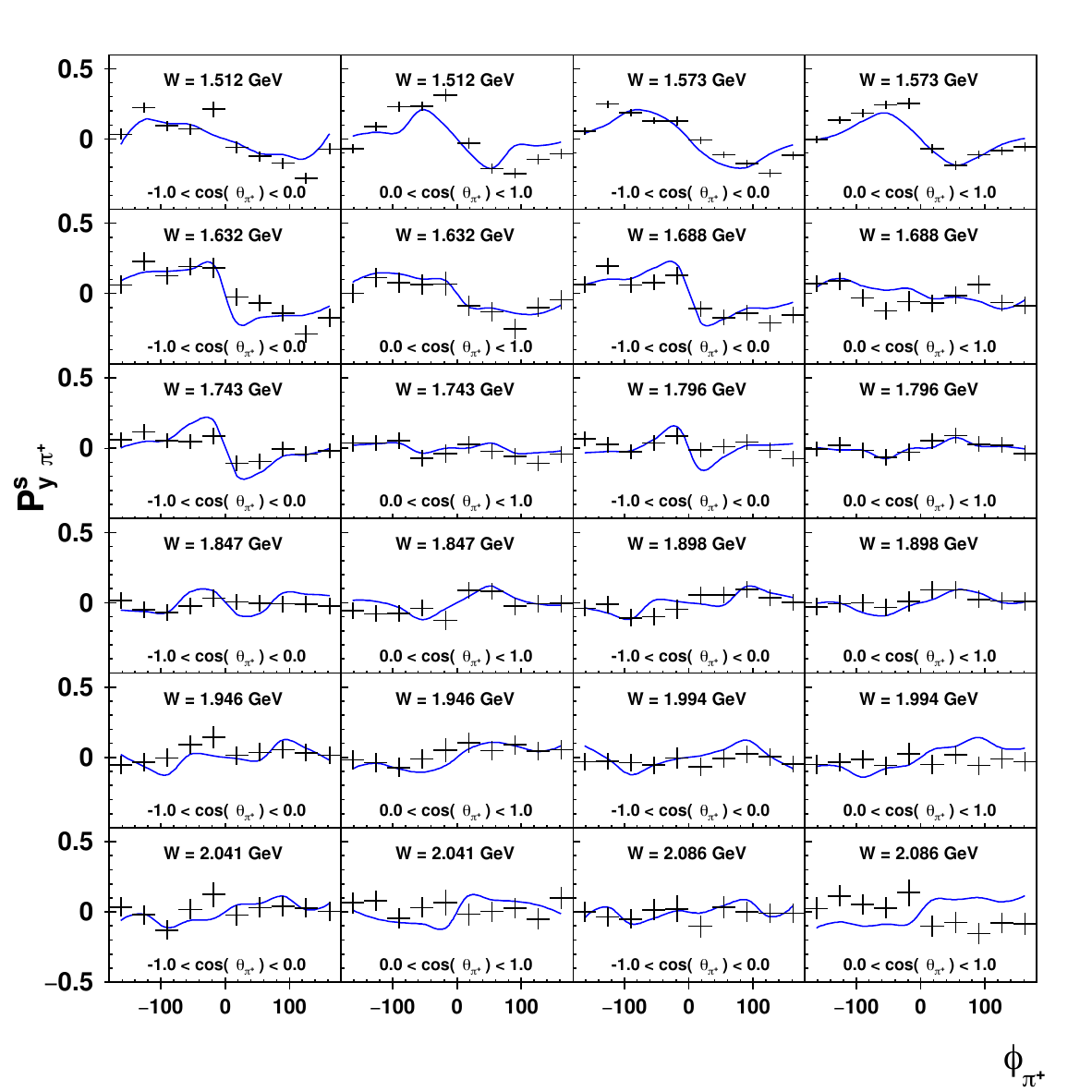}
    \caption{Full results for the beam-target observable, $\rm\bf P_y^s$, using a linearly polarized photon beam and
      a transversely polarized target in the reaction $\gamma p\to p\pi^+\pi^-$. The data have been analyzed for
      100-MeV-wide energy bins in the incident photon energy, $E_\gamma\in [700,1800]$~MeV, and are shown for
      bins in the corresponding center-of-mass energy~$W$. The data points are presented in the backward
      (cos\,$\theta < 0$) and forward direction (cos\,$\theta > 0$) are shown versus~$\phi$. These two quantities
      denote the polar and azimuthal angle of the $\pi^+$~meson in the rest frame (helicity frame) of the
      $\pi^+\pi^-$~system. The blue solid line denotes the BnGa-PWA solution.}\label{Figure:P_y_s}
  \end{center}
\end{figure*}

\begin{figure*}[t]
  \begin{center}
    \includegraphics[width=1.0\textwidth]{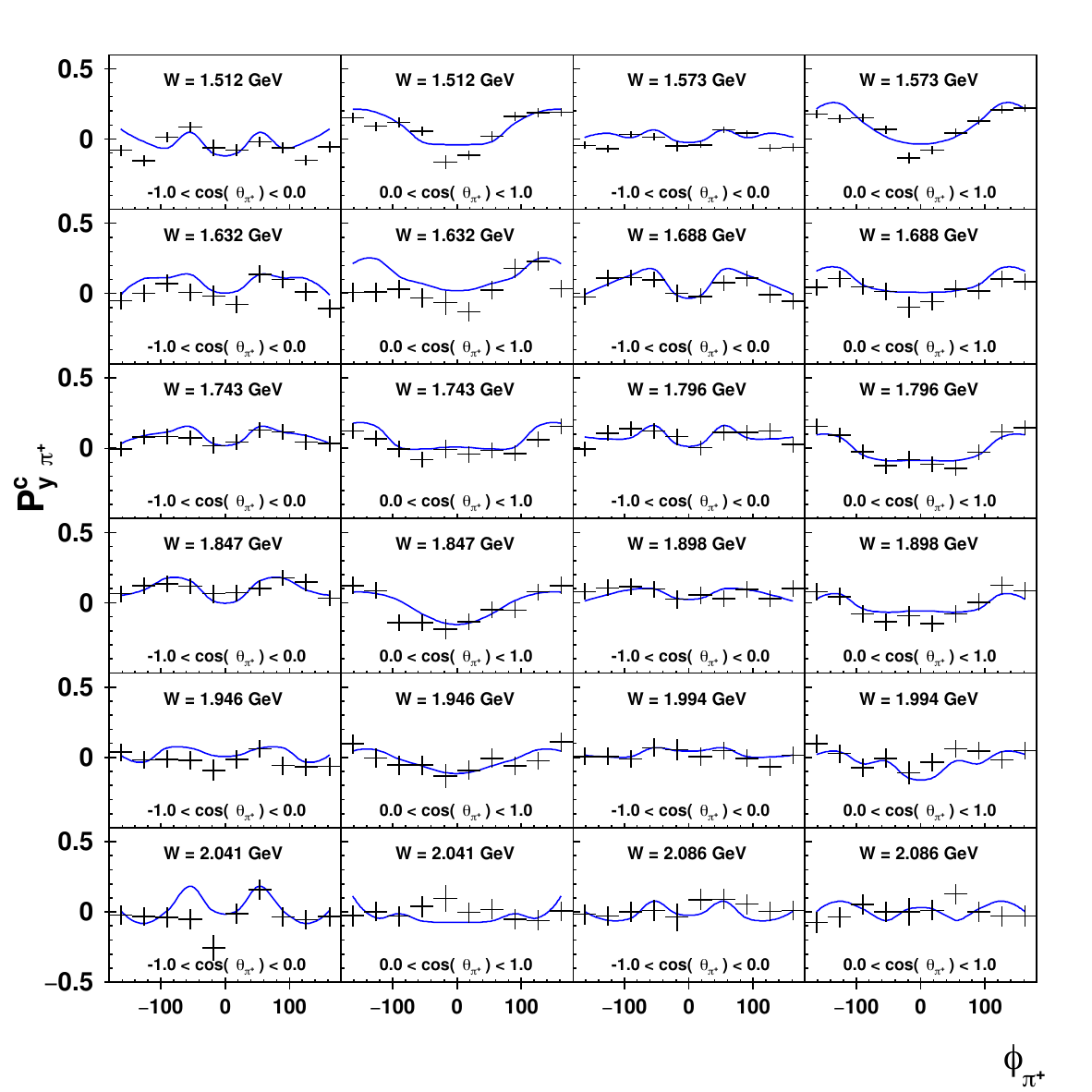}
    \caption{Full results for the beam-target observable, $\rm\bf P_y^c$, using a linearly polarized photon beam and
      a transversely polarized target in the reaction $\gamma p\to p\pi^+\pi^-$. The data have been analyzed for
      100-MeV-wide energy bins in the incident photon energy, $E_\gamma\in [700,1800]$~MeV, and are shown for
      bins in the corresponding center-of-mass energy~$W$. The data points are presented in the backward
      (cos\,$\theta < 0$) and forward direction (cos\,$\theta > 0$) are shown versus~$\phi$. These two quantities
      denote the polar and azimuthal angle of the $\pi^+$~meson in the rest frame (helicity frame) of the
      $\pi^+\pi^-$~system. The blue solid line denotes the BnGa-PWA solution.}\label{Figure:P_y_c}
  \end{center}
\end{figure*}

\subsection{The Single-Spin Asymmetries}
Figures~\ref{Figure:I_c}\,-\,\ref{Figure:I_s} show the CLAS-FROST results for the two beam asymmetries and
Figures~\ref{Figure:P_x}\,-\,\ref{Figure:P_y} the results for the two target asymmetries in the photoproduction reaction
$\gamma p\to p\pi^+\pi^-$ as the statistically weighted average of the Topologies~1,~2, and~4. The data points include
only the statistical uncertainties. The data have been analyzed for 100-MeV-wide energy bins in the incident photon
energy, $E_\gamma\in [700,1800]$~MeV, and are presented in the figures -- for better clarity -- for every other energy
bin using the corresponding center-of-mass energy. The $W$~value shown for each distribution refers to the center of
the corresponding $E_\gamma$~energy bin. The data points have been extracted in the rest frame of the
$\pi^+\pi^-$~system (helicity frame). The $\pi^+$~meson is chosen for the representation of the results and thus, the
angles cos\,$\theta$ and $\phi$ in the figures denote the polar and azimuthal angle of the $\pi^+$ in the helicity frame.
Sufficient statistics was available to bin the data in ten angle bins each. For a better overview, the results are again shown
for every other cos\,$\theta$~bin. The full set of experimental results is available as supplemental material. The
observables $\rm\bf I^s$ and $\rm\bf P_x$ exhibit the expected odd symmetry with respect to $\phi_{\pi^+} = 0$,
whereas the observables $\rm\bf I^c$ and $\rm\bf P_y$ show the expected even symmetry. The observables exhibit
rich structures and, in particular, the beam asymmetries acquire large values across a broad energy range.

\subsection{The Double-Spin Asymmetries}
Figures~\ref{Figure:P_x_s}\,-\,\ref{Figure:P_y_c} show the results for the four beam-target, double-polarization
observables in the photoproduction reaction $\gamma p\to p\pi^+\pi^-$ (Eq.~\ref{Equation:two_pion_reaction})
including only the statistical uncertainties for each data point. Given the limited statistics for the extraction of
double-spin asymmetries, the full set of results is presented. Ten data points are available for each distribution
in two cos\,$\theta_{\pi^+}$~angle bins: backward direction (cos\,$\theta_{\pi^+} < 0$) and forward direction
(cos\,$\theta_{\pi^+} > 0$). The observables $\rm\bf P_{x,y}^c$ exhibit the expected even symmtry with respect
to $\phi = 0$ and $\rm\bf P_{x,y}^s$ the expected odd symmetry. Overall, the double-polarization observables
acquire smaller values and are typically within $\pm 0.2$ or less.

\begin{table}[t!]
\caption{\label{tab:Sys_err} List of systematic uncertainties.}
\begin{center}
\begin{tabular}{lcc}
Source & & Systematic Uncertainty\\
\hline\hline
Background subtraction & ~~~ & 10\,\%\\
Beam-polarization & & $5\,\%$\\
Target-polarization & & $2\,\%$\\
Target-offset angle & & $2.5\,\%$\\
Normalization of data sets & &\\
\qquad beam asymmetries & & $5\,\%$\\
\qquad target asymmetries & & $2\,\%$\\
\qquad beam-target asymmetries & & $5\,\%$\\
\hline\hline
%Beam asymmetry & &\\
$\sigma_{\rm\,total}$ (fractional only) & & $\sim 13\,\%$
%Target asymmetry & &\\
%$\sigma_{\rm\,total}$ (fractional only) & & $\sim 3.5\,\%$\\
\end{tabular}
\end{center}
\end{table}

\subsection{Systematic uncertainties\label{ssec:sys_errors}}
The individual contributions to the overall systematic uncertainty for each observable that were studied in this analysis
are listed in Table~\ref{tab:Sys_err}. The fractional uncertainties were added in quadrature and the total is given in
Table~\ref{tab:Sys_err} and is approximately 13\,\%.

A major contribution came from the event-based background-subtraction technique. To estimate this contribution to
the overall systematic uncertainty, the $Q$~value of each event was increased by $\sigma_Q$ and the corresponding
asymmetry was re-extracted. Here, $\sigma_Q$ denotes the fit uncertainty in the $Q$~value of the $i^{th}$ event and
the uncertainty in the yield of any $\phi_{\rm lab}$~bin was thus equal to the sum over all $\sigma_{Q_i}$ within that bin.
Therefore, the method corresponds to the case of all events being 100\,\% correlated, which results in overestimated
uncertainties. An estimate of the overall fractional uncertainty is then based on:
\begin{align*}
  \begin{split}
    \left|\frac{\Delta x}{x}\right|_{av.} \, = \,\Biggl(~{\frac{\sum_{i=1}^\text{all 
    bins}{\Bigl\{\,\left|\frac{x \,-\, x'}{x}\right|\,\bigl(\frac{1}{\sigma_{x}}\bigr)^2\,\Bigr\}}}
    {\sum_{i=1}^\text{all bins}{\,\bigl(\frac{1}{\sigma_{x}}}\bigr)^2}}~\Biggr)\,,
 \end{split}
\end{align*}  
where $x$ is the value of the nominal polarization observable, and $x^{\,\prime}$ is the re-extracted polarization
observable and $\sigma_x$ is the statistical uncertainty in~$x$. 

The systematic uncertainty in the linear-beam polarization was evaluated to be~$\sim\,5\,\%$, a value which was also
used in other CLAS-FROST analyses~\cite{CLAS:2016wrl,CLAS:2013pcs}. The systematic uncertainty associated with
the target polarization was determined to be~$\sim\,2\,\%$~\cite{Keith:2012ad}. To estimate the systematic uncertainty
in the observable due to the target-offset angle, this angle was varied by its uncertainty of~$\pm 0.4^\circ$ and the
change in the re-extracted {\text{observable}} was examined. The effect was found to be $2\,\%$ on average.
 
For the study of the beam asymmetry, three factors were required to normalize the four linearly-polarized data
sets, as can be seen from Eqs.~\ref{equ:norm1}-\ref{equ:A_sigma_2} (Section~\ref{ssec:Likelihood}):
\begin{equation}
  N_1\,=\,\frac{\Phi^{+}_{\parallel}}{\Phi^{-}_{\parallel}}\,\bar{\Lambda}_R,\quad
  N_2\,=\,\frac{\Phi^{+}_{\perp}}{\Phi^{-}_{\perp}}\,\bar{\Lambda}_R,\quad
  N_3\,=\,\frac{\Phi^{+}_{\parallel}}{\Phi^{+}_{\perp}}~.
\end{equation}
The first two normalization factors were needed to {\it unpolarize} the target in the `$\parallel$' and `$\perp$' data sets,
respectively. The third normalization factor was then required to normalize the corresponding `$\parallel$' and `$\perp$'
data sets (after the target was rendered {\it unpolarized}). The uncertainties in the normalization factors depended on the
uncertainties in the flux ratios, which were obtained from the ratios of the numbers of reconstructed events originating
from the polyethylene target. One way to estimate the systematic uncertainty in these ratios was to compare them with
the ratios obtained from the carbon target. The results were found to differ by $2\,\%$ or less at all energies. Another
way to check the systematics of this method was to use the direct information on the photon flux from the photon
tagging system. Although this information was not available for the FROST data used in this analysis, it was available
for FROST-g9a data, which utilized a circularly-polarized beam and a longitudinally-polarized target. The results
differed again by only $\sim 2\,\%$ from those determined for the polyethylene target. The applied uncertainties
of~$2\,\%$ in the flux ratios as well as the uncertainty in the target polarization were used to evaluate the overall
uncertainties in the normalization factors using standard error propagation. Since each normalization factor could
be varied by $\pm\sigma$, all permutations were performed and the observable re-extracted. The change in the
asymmetries was observed to be $5\,\%$ on average across all energies.

All observables are shown as statistically weighted averages of the different topologies with the exception of the
missing-proton case (Topology~3). To understand possible acceptance effects from integrating over kinematic
variables in the five-dimensional phase space of this three-body final state, we studied the difference distributions
of each topology with respect to the corresponding average for every observable, and found these distributions
symmetric and centered at zero. The only exception was observed for the $\rm\bf I^{\,c}$~observable and the
missing-$\pi^+$ topology, which showed a small shift away from zero. In conclusion, acceptance effects are
considered extremely small and hardly separable from statistical fluctuations.

\section{Partial Wave Analysis\label{Section:PWA}}
The data presented here were included in a coupled-channel analysis within the Bonn-Gatchina (BnGa) PWA framework.
The BnGa~database takes into account almost all important data sets of photo- and pion-induced reactions, including
three-body final states~\cite{Anisovich:2011fc}. A full description of the experimental database~\cite{BnGa:Database}
goes beyond the scope of this paper. 

The angular distributions of the scattering amplitudes in the BnGa analysis for the production and the decay of baryon 
resonances are constructed in the framework of the spin-momentum operator expansion method. The details of this
method are discussed in Ref.~\cite{Anisovich:2006bc}. The dynamical part is represented by a modified $K$-matrix.
It incorporates nearly all nucleon and $\Delta$~resonances below~2.2~GeV reported in the
RPP~\cite{ParticleDataGroup:2022pth} and determines the decay fractions and transition residues to a large number of
final states. The inclusion of the data presented here and of other data on
$p\pi\pi$~\cite{CBELSATAPS:2015kka,A2:2022ipx,CBELSATAPS:2022uad, CLAS:2024iir} provides evidence for new
resonances, strengthens the evidence for the existence of some resonances with so far poor or fair evidence only,
and yields branching ratios for the decays of nucleon and $\Delta$~resonances into $N\rho$, $N\sigma$, and several
cascade decays via $\Delta(1232)\pi$ or several $N^\ast\pi$~\cite{Sarantsev:2025lik}.

\section{Summary}
Several single- and double-polarization observables have been extracted in $\pi^+\pi^-$~photoproduction off the
proton at Jefferson Laboratory using the CLAS spectrometer and the frozen-spin FROST target, covering the center-of-mass
energy range from 1.51 to 2.04~GeV. All presented observables are based on experiments using linear-beam and
transverse-target polarization, and are all first measurements for the $\pi\pi$~final state involving two charged
pions. These data and additional cross-section data from CLAS were included in a partial-wave analysis within the
Bonn-Gatchina framework. Resonance contributions and $N^\ast$~decays into $p\rho$ will be discussed in a
forthcoming publication.

\begin{acknowledgments}
The authors thank the technical staff at Jefferson Lab and at all the participating institutions for their invaluable
contributions to the success of the experiment. This material is based upon work supported by the U.S. Department
of Energy, Office of Science, Office of Nuclear Physics, under Contract No. DE-AC05-06OR23177. The group at Florida
State University acknowledges additional support from the U.S. Department of Energy, Office of Science, Office of
Nuclear Physics, under Contract No. DE-FG02-92ER40735. This work was also supported by the US National Science
Foundation, the State Committee of Science of Republic of Armenia, the Chilean Comisi\'{o}n Nacional de Investigaci\'{o}n
Cientifica y Tecnol\'{o}gica (CONICYT), the Italian Istituto Nazionale di Fisica Nucleare, the French Centre National de la
Recherche Scientifique, the French Commissariat a l'Energie Atomique, the Scottish Universities Physics Alliance (SUPA),
the United Kingdom's Science and Technology Facilities Council, and the National Research Foundation of Korea, the
Deutsche Forschungsgemeinschaft (SFB/TR110), and the Russian Science Foundation under grant 16-12-10267.
\end{acknowledgments}

\end{document}